\documentclass{JHEP3}
\usepackage{epsfig}
\usepackage{amssymb}

\newcommand{\be}{\begin{equation}}
\newcommand{\ee}{\end{equation}}
\newcommand{\bea}{\begin{eqnarray}}
\newcommand{\eea}{\end{eqnarray}}
\newcommand{\bean}{\begin{eqnarray*}}
\newcommand{\eean}{\end{eqnarray*}}
\newcommand{\mat}[1]{\left( \matrix{#1} \right)}
\newcommand{\tmat}[1]{{\scriptsize \mat{#1}}}

\newcommand{\myhref}[1]{\href{http://xxx.arXiv.gov/abs/#1}{#1}}

\def\IZ{\mathbb{Z}}

\def\IC{\mathbb{C}}
\def\IP{\mathbb{P}}

\def\beq{\begin{equation}}
\def\eeq{\end{equation}}

\def\Tr{\mathop{\rm Tr}}
\newcommand{\fref}[1]{Figure~\ref{#1}}
\def\eref#1{(\ref{#1})}
\newcommand{\sref}[1]{\S\ref{#1}}

\def\nn{\nonumber}

\preprint{MIT-CTP-3386\\UPR-1046-T\\ {\tt hep-th/0306092}}

\title{Duality Walls, Duality Trees and Fractional Branes}

\author{Sebastian Franco$^1$, Amihay Hanany$^1$, Yang-Hui He$^2$ and
Pavlos Kazakopoulos$^1$
\\
1. Center for Theoretical Physics,
\\Massachusetts Institute of Technology,\\ Cambridge, MA 02139, USA\\ 
~\\
2. Department of Physics,\\
The University of Pennsylvania,\\
209, S.~33rd st.,\\
Philadelphia, PA 19104-6396\\
~\\
\email{sfranco,hanany,noablake@mit.edu,yanghe@physics.upenn.edu}
}

\abstract{We compute the NSVZ beta functions for ${\cal N}=1$
four-dimensional quiver theories arising from D-brane probes on
singularities, complete with anomalous dimensions, for a large set of
phases in the corresponding duality tree.
While these beta functions are zero for D-brane probes, they are
non-zero in the presence of fractional branes. As a result there is a
non-trivial RG behavior. We apply this 
running of gauge couplings to some toric singularities such as the
cones over Hirzebruch and del Pezzo surfaces. We observe
the emergence in string theory, of ``Duality Walls,'' a finite energy
scale at which the number of degrees of freedom becomes infinite, and beyond
which Seiberg duality does not proceed. We also identify certain
quiver symmetries as T-duality-like actions in the dual holographic
theory.}
\begin{document}

\section{Introduction}

The emergence of the Klebanov-Strassler ``Cascade Phenomenon''
\cite{klst} has
been a marvelous motif in the grand theme of the AdS/CFT
correspondence. In particular, the study of the relation between the
supergravity bulk theory and the field theory on world-volumes of
branes probing singular geometries has been part of the
on-going programme to extend the correspondence to more general, for
example non-conformal, classes of field theories.

Indeed, with the introduction of $M$ fractional D3 branes in addition to
$N$ regular ones on our familiar conifold singularity, \cite{klst}
studied the dual field theory which is a 4-dimensional ${\cal N}=1$
$SU(N+M) \times SU(N)$ non-conformal gauge theory. 
The radial variation of the
fiveform flux in AdS is identified with the running of the gauge
couplings. Therefore when one of the couplings becomes strong, one
can, \`{a} la Seiberg, dualise the theory and flow to the IR,
to one with gauge group $SU(N) \times SU(N-M)$.
One may follow this RG flow in the field
theory in principle {\it ad infinitum} and the process was referred to
as a cascade \cite{klst}.

The generalisation of this phenomenon to other geometries is hindered
by the fact that the conifold is really the only geometry for which we
know the metric. Nevertheless nice extensions from the field theory
side have been performed. Notably, in \cite{fiol}, the cascade has been
recast into properties of the Cartan matrix of the quiver
\cite{kac,review}. Then Seiberg duality becomes Weyl reflections in
the associated root space. The UV behaviour would thus depend markedly
on whether the Cartan matrix is hyperbolic (with a single negative
eigenvalue and the rest positive) or not. Indeed for some simple quiver
examples without consideration for stringy realisation,
it was shown that the RG flow converges in the UV and
surprisingly there is a 
finite accumulation point at which the scale of the
dualisations pile up. This casts a shadow of doubt as to possible UV
completions of these field theories. Such fundamental limitation on
the scale of the theory was originally dubbed ``Duality Walls'' in
\cite{strassler}.

The issue seems to persist as one studies generalisations of the
conifold geometry and in realisations in string theory. As a first
example that is chiral and arising from standard string theory
constructions, \cite{HW} discussed the case of our familiar $\IC^3 /
\IZ_3$ singularity. Using na\"{\i}ve beta-functions without
consideration for the anomalous dimensions, \cite{HW} analysed in
detail how one encounters duality walls for this string theoretic
gauge theory.

Indeed, despite our present lack for metrics and supergravity
solutions for wider classes of examples, we are in fact well armed
from the gauge theory perspective. Extensive methodology and
catalogues of non-spherical horizons have been in circulation 
(q.v.~e.g.~\cite{Morrison,Bergman,Gubser,Herzog,domo,HH,BGLP}). 
A particular class for which an algorithmic outlook was
partaken is the toric singularities of which the conifold and the $\IC^3 /
\IZ_3$ orbifold are examples 
\cite{BGLP,toric,fh2}. For these geometries,
Seiberg-like dualities dubbed ``Toric Duality'' \cite{toric,fhhu,BP}
have been labouriously
investigated. Consecutive application of such a duality on a given
theory essentially translates to a systematic application of certain
quiver transformation rules. These rules can be understood from many
fruitful perspectives: as ambiguities in the Inverse Toric Algorithm
\cite{toric};
or as monodromy transformations of wrapped cycles around
the singularity \cite{ceva,hiv,fhhi,cfikv}; 
or as braiding relations in $(p,q)$ sevenbrane
configurations \cite{fhhi}; 
or as mutations in helices of exceptional collections
of coherent sheafs \cite{hiv}; 
or as tilting functors in the D-brane derived
category \cite{bedo,volker}, etc.

A key feature of such a duality is the {\bf tree structure} of the
space of dual theories. As we dualise upon a node in the quiver at
each stage, a new branch blossoms. The topology of the tree is
important. For example, whether there are any closed cycles which
would signify that certain dualities may be trapped within a group of
theories.

We are therefore naturally inspired by the conjunction of the toy
model in \cite{HW} and our host of techniques from toric duality.
Dualisations in the tree is precisely the desired cascading
procedure.
A first care which needs to be taken is a thorough analysis of the
beta-function, including the anomalous dimensions. Happily, the form
of the exact beta function with the anomalous dimensions
has been computed by \cite{NSVZinstanton} 
for ${\cal N}=1$ gauge theories and
by \cite{LS} for quiver theories in particular. 

In ${\cal N}=1$ SCFT theories, the 't Hooft anomaly for the
$U(1)_R$ charge determines 
all anomalous dimensions of chiral operators. These however are 
determined up to global non-R flavour symmetries. Computationally one
must resort to finding such additional symmetries. In the
quiver cases, these can be done by guided inspection of the quiver
diagrams \cite{symmetries}.

Recently, the nice
work by \cite{Intri1} posited a maximisation principle to
systematically determine the R-symmetry and hence all
anomalous dimensions.
Namely one must maximise a certain combination of the
$U(1)$ charges. This quantity is called $a$ in the canonical
literature and together with $c$ constitute the central charges of the 
SCFT. Indeed it is believed that $a$ obeys a 4d version of
Zamolodchikov's $c$-theorem , having a monotone increase along RG flow
toward the UV. The central 
charge $a$  
shall be for us, a measure of the number of degrees of freedom in the
field theory. 
In the AdS dual, we
can view this as the thermodynamic entropy of the horizon
of the singular geometry.


The structure of this paper is as follows.
We will first require three ingredients the combination of
which will form the crux of our calculation. The first piece we need
is four dimensional ${\cal N}=1$ super
conformal field theory (SCFT), especially quiver theories. 
In particular we remind the reader of the
computations of anomalous dimensions in the beta function. This will
be the subject of \sref{section_computing_gammas}.
The second piece we need is the so-called ``duality trees'' which
arise from iterative Seiberg-like dualisations of quiver
theories. This, with concrete examples from the zeroth del Pezzo, 
will constitute \sref{section_trees}.
The final piece we need is to recall the rudiments of the
Klebanov-Strassler ``cascade'' for the quiver theory associated to the
conifold. We do this in \sref{section_conifold_cascade}.

Thus equipped, we examine a simple but illustrative gauge theory in
\sref{section_F0}. This is a fairly well-studied 
quiver theory arising from D3-branes probing the singular complex cone
over the zeroth Hirzebruch surface. The duality tree for the conformal
phases of the theory form a flower.
With the appropriate addition of
fractional branes to take us away from conformality, we compute the
beta function running in \sref{section_closed_cycle}, 
by determining, using the
abovementioned maximisation principle, all anomalous dimensions. 
We will find in \sref{section_wall}
that there is indeed a duality wall, viz., an energy
scale beyond which dualities cannot proceed. Interestingly, certain
quiver automorphism symmetries can be identified with T-duality-like
actions in the
dual AdS theory.

Such analyses are well adapted and easily generalisable to arbitrary
quiver theories. As a parting example, we present the case of the cone
over the first del Pezzo surface in \sref{sec:dP1}. This case exhibits
another interesting phenomenon which we call ``toric islands.''
We end with concluding remarks and prospects in
\sref{sec:con}. Moreover, in Appendix 1, we use the method of
Picard-Lefshetz monodromy transformations to analytically determine
the asymptotic behaviour of some of the cascades, especially in the case
of alternating dualisations. Finally, as mentioned above, the central
charge 
$a$ actually measures the entropy and hence the volume of base of the AdS
dual geometry. As an application, we compute catalogue in Appendix 2,
various horizon volumes, particularly the Abelian orbifold
$\IC^3/\IZ_n$ and the cones over the del Pezzo surfaces.

\section{Computing Anomalous Dimensions in a SCFT}
\label{section_computing_gammas}

We devote this section to a summary of beta-functions
in 4D ${\cal N}=1$ SCFT and
of how to compute in particular the anomalous dimensions. The
remainder of the paper will make extensive use of the values of these
anomalous dimensions.

The necessary and sufficient conditions for a ${\cal N}=1$
supersymmetric gauge theory with superpotential to  
be conformally invariant, i.e., a SCFT, are (1) the vanishing of the
beta function for 
each gauge coupling and (2) the requirement 
that the couplings in the superpotential be dimensionless. 
Both these conditions impose constraints on the 
anomalous dimensions of the matter fields, that is, chiral operators
of the theory. This is because supersymmetry relates
the gauge coupling beta functions to the anomalous dimensions of the
matter fields due to the form of the
Novikov-Shifman-Vainshtein-Zakharov(NSVZ) beta functions.
\cite{NSVZinstanton,LS}.

The examples which we study in this paper are a class of SUSY gauge
theories known as quiver theories.
These have product gauge groups of the form $\prod\limits_i U(N_{c_i})$
together with $N_{f_i}$ bifundamental 
matter fields for the $i$-th gauge factor. There is also a 
polynomial superpotential. Such theories can be conveniently encoded
by quiver diagrams where nodes are gauge factors and arrows are
bifundamentals \cite{domo,review}.
The SCFT conditions for these quiver theories can be written as:
\bea
\beta_i \sim 3N_{c_i}-N_{f_i}+\frac{1}{2}\sum_j{\gamma_j}=0  \nn \\
-d(h)+ \frac{1}{2}\sum_k{\gamma_k}=0 
\label{betas}
\eea
where $\beta_i$ is the beta-function for the $i$-th gauge factor,
and $\gamma_j$, the anomalous dimensions.
The index $j$ labels the fields charged under the $j$-th gauge group
factor while $k$ indexes the fields appearing in the $k$-th term of
the superpotential with coupling $h$, whose na\"{\i}ve mass dimension
is $d(h)$.

These conditions \eref{betas} constitute a linear system of 
equations. However they do not always uniquely determine the anomalous
dimensions because there will be more variables than constraints. One
or more of the $\gamma$'s are left as free parameters. 
Recently, Intriligator and Wecht \cite{Intri1} provided a general
method for 
fixing this freedom in aribitrary 4D SCFT, 
whereby completely specifying the anomalous
dimensions. They showed that the R-charges of the matter fields, which
in an SCFT are related to the $\gamma$'s,
are those that (locally) maximize the central charge $a$ of the 
theory. The central charge $a$ is given in terms of the R-charges by
\beq
a=\frac{3}{32}(3\Tr R^3-\Tr R) \ ,
\label{definition_a_1}
\eeq
where the trace is taken over the fermionic components of the vector
and chiral multiplets.  

In a quiver theory with gauge group $\prod\limits_i U(N_i)$ 
and chiral bifundamental multiplets with multiplicities $f_{ij}$
between the $i$-th and $j$-th gauge factors (the matrix $f_{ij}$ is
the adjacency matrix of the quiver), we can give an explicit 
expression for \eref{definition_a_1} in terms of the R-charges $R_{ij}$
of the lowest components of the bifundamentals
\beq
a=\frac{3}{32} \left[2 \sum_i N_i^2 + \sum_{i<j} f_{ij} N_i N_j \left[
3 (R_{ij}-1)^3 - (R_{ij}-1) \right ]\right] \ .
\label{definition_a_2}
\eeq
Parenthetically, we remark that in some cases, such as the ones to be
discussed in \sref{section_conifold_cascade}  and
\sref{section_F0}, anomalous dimensions can be fixed by using some
discrete symmetries by inspecting the quiver 
and the form of the superpotential, without the need to appeal to the
systematic maximization of $a$.

Now in \eref{definition_a_2} we need to know the R-charges. However,
in a SCFT the conformal dimension $D$ of a chiral operator is related
to its R-charge by $D=\frac{3}{2}|R|$. Moreover, for bifundamental  
matter the relation between $D$ and $\gamma$ is
$D=1+\frac{\gamma}{2}$. Therefore we can write the R-charges and hence
\eref{definition_a_2} in terms of the anomalous dimensions by 
\beq
\frac{3}{2}|R| = 1+\frac{\gamma}{2} \ .
\eeq
Therefore, after solving the conformality constraints \eref{betas} 
we can write $a$ in terms of the
still unspecified $\gamma$'s by \eref{definition_a_2} and then
maximize it in order to completely determine all the anomalous
dimensions. 

The freedom in the anomalous dimensions after using the SCFT
conditions reflects the presence of non-anomalous $U(1)$ flavor
symmetries in the IR theory. Initially, there is one $U(1)$
flavor symmetry for each arrow of the quiver. 
All the matter fields lying on an arrow have the same charge under
this $U(1)$. Now we must impose the anomaly free condition for each
node, this is the condition that for the adjacency matrix $f_{ij}$ 
at the $i$-th node we have
\beq
\sum_j f_{ij} N_j = \sum_j f_{ji} N_j \ .
\eeq
In other words, the ranks of the gauge groups, as a vector, must lie
in the integer nullspace of the antisymmetrised adjacency matrix:
\beq
\label{anofree}
(f - f^T)_{ij} \cdot \vec{N} = 0 \ .
\eeq
Indeed the matrix $(f - f^T)$ is the intersection matrix of the quiver
in geometrical engineering of these theories
(q.v.~e.g. \cite{hi,hiv,fhhi}). 
After imposing this condition \eref{anofree}, 
we are left with ($\#$ of arrows - $\#$ of nodes)
non-anomalous $U(1)$'s. 
The invariance of the superpotential reduces their 
number even more, giving one linear relation between their charges for
each of its terms. But the number 
of independent such relations is not always sufficient to eliminate
all the abelian flavor symmetries. The  
charges of those that still remain in the IR can indeed be read off
from the expressions for the anomalous dimensions  
of the fields in terms of those that remain free after imposing the
conditions \eref{betas}. 

The way in which the charge matrix of the remaining $U(1)$
flavor symmetries appears in this framework is as the matrix
of coefficients that express the anomalous dimensions of
the bifundamental fields as linear combinations of numerical
constants and some set of independent anomalous dimensions.
Specifically, suppose we start with $n$ anomalous dimensions and
that the solution to \eref{betas} specifies $k$ of them in terms of
the other $n-k$:

\beq
\gamma_i=\gamma_{0_i}+q_{ij}\gamma_j  \ \ , \ \  i=1,\ldots,k \ \ ,\ \
j=k+1,\ldots,n . 
\label{red}
\eeq

The corresponding R-charges are related to the $\gamma$'s by
$R=\frac{\gamma+2}{3}$. The charges of the matter fields under
these residual $U(1)$'s from are given by the $q_{ij}$
matrix in (\ref{red}). The constants $\gamma_{0_i}$ are mapped
to the test values of R charges. It is important to keep in mind
that it is possible to change the basis of $U(1)$'s (correspondingly
the set of independent anomalous dimensions), in which case
the charge matrix would be modified.
\section{Duality Structure of SUSY Gauge Theories: Duality Trees}

\label{section_trees}

Having reminded ourselves of the methodology of computing anomalous
dimensions, we turn to the next ingredient which will prepare us for
the cascade phenomenon, viz., the duality trees
which arise from Seiberg-like dualities performed on the quivers.
An interesting way to encode dual gauge theories and their relations
is by using {\bf duality trees}. This construction was introduced in
\cite{cfikv}, for the specific case of D3-branes probing a complex
cone over $dP_0$, the zeroth del Pezzo surface.

In general, for a quiver theory with adjacency matrix $f_{ij}$ and 
$n$ gauge group factors, there
are $n$ different choices of nodes on which to perform Seiberg
duality.  In other words, we can dualise any of the $n$ nodes of the
quiver to obtain a new one, for which we again have $n$ choices for
dualisation. We recall that dualisation on node $i_0$ proceeds as
follows. Define $I_{in} :=$ nodes having arrows going into $i_0$, 
$I_{out} := $ those having arrow coming from $i_0$ and
$I_{no} :=$ those unconnected with $i_0$.
\begin{enumerate}
\item   Change the rank of the node $i_0$ from $N_c$ to $N_f-N_c$
	with $N_f=\sum\limits_{i\in 
	I_{in}} f_{i,i_0} N_i = \sum\limits_{i\in I_{out}} f_{i_0,i}
	N_i$;
\item   $f^{dual}_{ij} = f_{ji} \qquad \mbox{ if either }i,j = i_0$;
\item   Only arrows linking $I_{in}$ to $I_{out}$ will be changed and
        all others remain unaffected;
\item	$f^{dual}_{AB} = f_{AB} - f_{i_0 A} f_{B i_0} \qquad \mbox{ for
        } A \in I_{out},~~B \in I_{in}$;\\
        If this quantity is negative, we simply take it to mean 
        an arrow going from $B$ to $A$. This step is simply the
	addition of the Seiberg dual mesons (as a mass deformation if
	necessary).
\end{enumerate}
We remark that the fourth of these dualisation rules accounts for 
the antisymmetric part of the intersection matrix, which does not
encode bi-directional arrows. Such subtle cases arise when there are
no cubic superpotentials needed to give masses to the fields
associated with the bi-directional arrows. 
We encountered such a situation in \cite{symmetries}.

The subsequent data structure is that of a tree, where
each site represents a gauge theory, with $n$ branches emanating
therefrom, connecting it to its dual theories. This is called a
``duality tree.''
We will see along the paper that duality trees exhibit an extremely
rich  structure, with completely distinct topologies for the branches
for gauge theories
coming from different geometries.

\begin{figure}[ht]
  \epsfxsize = 3.5cm
  \centerline{\epsfbox{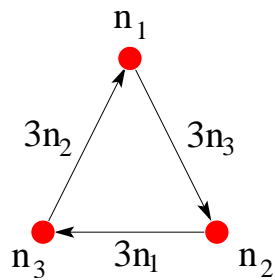}}
  \caption{Generic quiver for any of the Seiberg dual theories in the
  duality tree 
  corresponding to a D3-brane probing $\IC^3/\IZ_3$, the complex cone
  over $dP_0$.}
  \label{quiver_dP0}
\end{figure}

As an introduction, let us recall the simple example considered in
\cite{cfikv, fhhi}. The probed geometry in this case was a complex cone  
over $dP_0$. This cone is simply the famous non-compact $\IC^3/\IZ_3$
orbifold singularity.
The generic quiver for any one in the tree of Seiberg dual theories 
for this geometry will have the form as given in \fref{quiver_dP0}. 
The superpotential is cubic because there are only cubic gauge
invariant operators in this theory, given by closed loops in the quiver
diagram.

\begin{figure}[ht]
  \epsfxsize = 9cm
  \centerline{\epsfbox{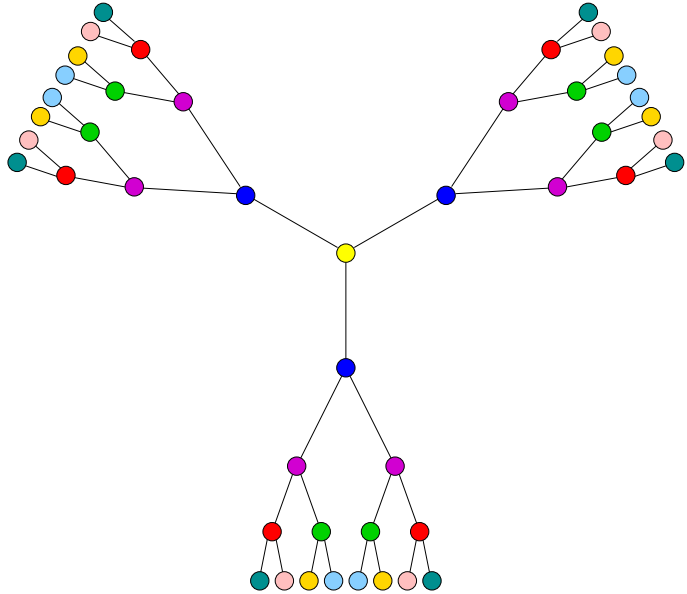}}
  \caption{Tree of Seiberg dual theories for $dP_0$. Each site of the
  tree represents a gauge theory, and the branches between sites 
  indicate how different theories are related by Seiberg duality
  transformations.}
  \label{dP0_tree}
\end{figure}

\begin{figure}[ht]
  \epsfxsize = 17cm
  \centerline{\epsfbox{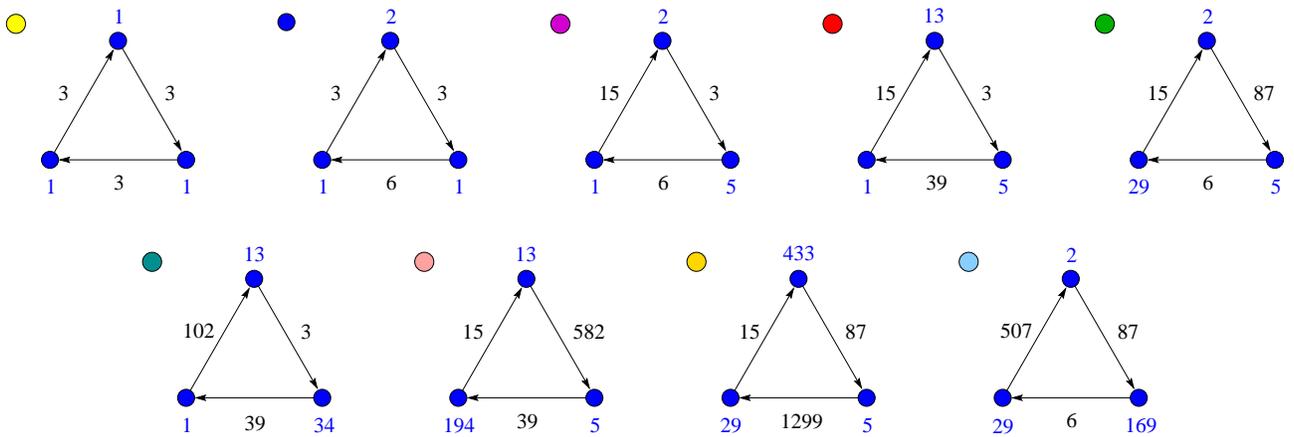}}
  \caption{Some first cases of the Seiberg dual phases in the duality 
  duality tree for the theory
  corresponding to a D3-brane probing $\IC^3/\IZ_3$, the complex cone
  over $dP_0$.}
  \label{dP0_references}
\end{figure}

Since there are three gauge group factors, there will be three branches
coming out from each site in the duality tree.  
The tree is presented in \fref{dP0_tree}. For clarity we colour-coded
the tree so that sites of the same colour correspond to equivalent
theories, i.e., theories related to one another 
by some permutation of the gauge groups and/or charge conjugation of
all fields in the quiver (in other words theories whose quivers are
permutations and/or transpositions of each other). We have also
included, the quivers to which the various coloured sites correspond
in \fref{dP0_references}.

One important invariant associated to an algebraic singularity is the
trace of the total monodromy matrix around the singular point. 
This can typically be recast into an
associated Diophantine equation in the intersection numbers, i.e., the
$f_{ij}$'s \cite{ceva,DeWolfe:1998eu,fhhi}. This equation captures all
the theories
that can be obtained by Seiberg duality and hence classifies the sites
in the tree. We remind the reader of the relation between this
equation and Picard-Lefschetz monodromy transformations in Appendix 1
and refer him/her to e.g.~\cite{fhhi}. 

From \fref{quiver_dP0}, we see that
there exists a simple relation between the intersection numbers and
the ranks of the gauge groups for $dP_0$, namely for rank
$(n_1,n_2,n_3)$, the intersection matrix is given by
$3 \tmat{
0 & n_1 & -n_3 \cr -n_1 & 0 & n_2 \cr n_3 & -n_2 & 0 \cr
}$. The Diophantine equation in terms of the ranks reads
\beq
n_1^2+n_2^2+n_3^2-3 n_1 n_2 n_3=0 \ .
\label{diophantine}
\eeq
This turns out to be the well-studied Markov equation.

It is important to stress that, up to this point, duality trees do not
provide any information regarding RG flows. In fact, if 
the theories under study are conformal the trees just represent the
set of dual gauge theories and how they are  
interconnected by Seiberg duality transformations within the conformal
window. 
We will extend our discussion about this point in
\sref{section_conifold_cascade} and \sref{section_F0}, 
where we will obtain non-conformal theories by the inclusion of
fractional branes.

\section{The Conifold Cascade}

\label{section_conifold_cascade}

A famous example of successive Seiberg dualisations is the
Klebanov-Strassler cascade in gauge theory \cite{klst} associated to
the warped deformed conifold \cite{klst}. In light of the duality tree
structure in the previous section, we now present the third and last
piece of preparatory work and summarise some key features
of this example, in order to illustrate the concept of duality
cascade, as well as to introduce  
many of the methods and approximations that will be used later.

\begin{figure}[ht]
  \epsfxsize = 4cm
  \centerline{\epsfbox{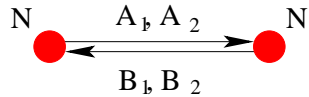}}
  \caption{Quiver diagram for the gauge theory on $N$ D3-branes
  probing the conifold.}
  \label{quiver_conifold}
\end{figure}

Let us begin by considering the gauge theory that appears on a stack
of $N$ D3-branes probing 
the conifold. This theory has an $SU(N) \times SU(N)$ gauge
symmetry. The matter content consists of 
four bifundamental chiral multiplets $A_{1,2}$ and $B_{1,2}$
and the quiver diagram is shown in \fref{quiver_conifold}. 
This model has also interactions given by the following quartic
superpotential
\beq
W={\lambda \over 2} \epsilon^{ij}\epsilon^{kl}\Tr A_i B_k A_j B_l
\eeq
for some coupling $\lambda$, and where we trace over color indices.

\begin{figure}[ht]
  \epsfxsize = .7cm
  \centerline{\epsfbox{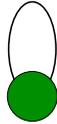}}
  \caption{The ``duality tree'' of the conifold. 
	Its single site represents the standard
	$SU(N) \times SU(N)$ theory.
	The closed link coming out the site and returning to
	it represents the fact that the theory, being self-dual, 
	transforms into itself under Seiberg  duality.}
  \label{tree_conifold}
\end{figure}

This gauge theory is self dual under Seiberg duality
transformations, by applying the duality rules in \sref{section_trees}. 
Accordingly, its ``duality tree'' is the 
simplest one, consisting of a single point representing the 
$SU(N) \times SU(N)$ theory, which transforms into  
itself when dualizing either of its two gauge groups. This is shown in 
\fref{tree_conifold}.
\bigskip

We will see below that 
when we apply the procedure for finding anomalous dimensions
outlined in \sref{section_trees} to
this specific case, taking into account  
its symmetries, we conclude that all the anomalous dimensions are in
fact equal to $-1/2$ and that the theory is conformal, 
i.e. both the gauge and superpotential couplings have vanishing beta
functions and \eref{betas} are satisfied.
In order to induce a non-trivial RG flow the theory has to be
deformed. A possible way of doing this 
is by the inclusion of fractional branes \cite{klst}. It is
straightforward to see what kind of fractional 
branes can be introduced. 

In general, introducing fractional branes is done by determining, for
a given quiver, the most general gauge groups consistent with anomaly
cancellation. Now recall from \eref{anofree}, possible ranks of the
gauge factors must reside in the integer nullspace of the intersection
matrix. Therefore a basis for probe
and fractional branes is simply given by a basis for this nullspace.
For the conifold, we find that the most general 
gauge group is $SU(N+M) \times SU(N)$. We will refer to $N$ as the
number of probe branes and to $M$ as the 
number of fractional branes. 

We see that indeed, for any non-vanishing $M$, there  
is no possible choice of anomalous dimensions satisfying \eref{betas}
and thus we are indeed moving away from the conformal point.
This case has been widely studied 
(see \cite{klst,peter} and references therein) and leads to a {\bf duality
cascade}. What this means is that at every step in the dualisation
procedure of this now non-conformal quiver theory, one of  
the gauge couplings is UV free while the other one is IR free. As we
follow the RG flow to the IR, we reach a  
scale at which the inverse coupling of the UV free gauge factor
vanishes. At this point, it is convenient to  
switch to a more suitable description of the physics, in terms of
different microscopic degrees of freedom,  
by performing a Seiberg duality transformation on the strongly coupled
gauge group. This procedure generates
the duality cascade when iterated.
Indeed, the tree of \fref{tree_conifold} can be interpreted as
representing a 
duality cascade modulo fractional brane contributions.

\subsection{Moving Away from the Conformal Point}
Let us now study this cascade in detail, setting the framework we will
later use to analyze cascades for general quiver theories. 
Recall, from \eref{betas}, that a key ingredient required for the
computation of the beta functions are the values of the anomalous
dimensions. We have already provided a method to compute anomalous 
dimensions in the absence of fractional branes, that is, in a
conformal theory, in \sref{section_computing_gammas}. 
There is no analogue for such a procedure when the theory is taken
away from conformality. However it is possible to work in a limit such
that their values are under control and the beta functions that govern
the RG flow can be computed to some approximation. Since 
$M/N$ measures the departure from conformal invariance, any anomalous
dimension will be of the generic form 
\beq
\gamma=\gamma_c+O(M/N)
\label{expansion_gamma1} \ ,
\eeq
where $\gamma_c$ is its value for the conformal case of $M=0$.
Now, this theory is symmetric under the transformation
\begin{eqnarray}
&&M \rightarrow -M \\ \nonumber
&&N \rightarrow N+M \ ,
\label{symmetry_conifold}
\end{eqnarray}
which, in the limit $N/M  \ll 1$ (i.e., we are taking a standard large
$N$ limit), simplifies to
\begin{eqnarray}
&&M \rightarrow -M \\ \nonumber
&&N \rightarrow N \ .
\end{eqnarray}
This indicates that, in fact, at large $N$,
$\gamma$ must be an even function in $M/N$
and so the expansion \eref{expansion_gamma1} has to start from the
second order \cite{klst}:
\beq
\gamma=\gamma_c+O(M/N)^2 \ .
\label{expansion_gamma2}
\eeq

The expression \eref{expansion_gamma2} is of great aid to us as it 
gives us the control over the anomalous dimensions we were
pursuing. Inspecting \eref{betas},
we see that because the departure of the $\gamma$'s from their
conformal values is of order $(M/N)^2$ at large $N$, 
the order $(M/N)$ contributions to the beta functions can be computed
simply by substituting the anomalous dimensions calculated at the
conformal point into \eref{betas}, and using the gauge groups with
the $M$ corrections. 

Let us be concrete and proceed to compute the cascade for this
example. First let us consider the anomalous dimensions
at the conformal point where $M=0$. They are the result of requiring
the beta functions for both $SU(N)$ 
gauge groups and for the single independent coupling in the
superpotential to vanish in accordance with \eref{betas}.  
In this case, these three conditions coincide and are reduced to
\beq
\gamma_{c,A}+\gamma_{c,B}=-1 \ ,
\eeq
where $\gamma_{c,A}$ (resp.~$\gamma_{c,B}$)
 is the critical value for the anomalous dimension for field $A$
(resp.~$B$).
Once we take into account the symmetry condition
$\gamma_{c,A}=\gamma_{c,B}$, we finally obtain 
\beq
\gamma_c=-1/2 \ .
\eeq 

Now let us consider the beta functions for the gauge couplings in the
non-conformal case of $M \neq 0$. They are
\beq
\begin{array}{rcl}
SU(N+M): & \ \ \ & \beta_{g_1}=N(1+\gamma_A+\gamma_B)+3M \\
SU(N):   & \ \ \ &
	\beta_{g_2}=N(1+\gamma_A+\gamma_B)+(-2+\gamma_A+\gamma_B)M.
\label{betas_conifold1}
\end{array}
\eeq
Note that there is no solution to the vanishing of these beta functions for $M\not =0$.
Replacing the anomalous dimensions at the conformal point 
$\gamma_c=-1/2$ into \eref{betas_conifold1} we obtain the 
leading contribution to the beta functions
\beq
\begin{array}{rcl}
SU(N+M): & \ \ \ & \beta_{g_1}=+3M \\
SU(N):   & \ \ \ & \beta_{g_2}=-3M \ .
\label{betas_conifold2}
\end{array}
\eeq

\begin{figure}[ht]
  \epsfxsize = 8cm
  \centerline{\epsfbox{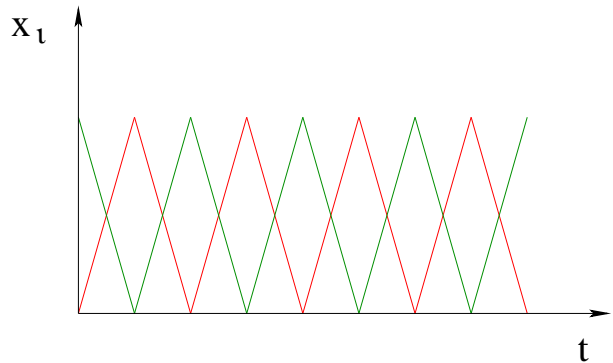}}
  \caption{Running of the inverse square gauge couplings $x_{i} =
  \frac{1}{g_{i}^2}$, i=1,2. against the log of energy scale $t = \log \mu$,
  for the conifold. The
  distance between consecutive dualizations is constant  
  and the ranks of the gauge groups grow linearly with the step in the
  cascade.}
  \label{couplings_conifold}
\end{figure}

Since the theory at any point in the cascade is given by the quiver in
\fref{quiver_conifold} with gauge group
replaced by $SU(N+(n+1)M) \times SU(N+n M)$ for some $n \in \IZ$ 
(where the role of the two
gauge groups is permuted at every step),
we see that the gauge couplings run as shown in
\fref{couplings_conifold},  where the beta function for each gauge
group changes from $\pm 3M$ to $\mp 3M$ with each dualization. 
In \fref{couplings_conifold} we use he standard notation to which we
adhere throughout the paper: the squared inverse couplings are denoted
as $x_i=1/g_i^2$ and the logarithm of the scale is $t=\log \mu$. 

An important feature of this RG flow is that the separation
between successive dualizations in the $t$ axis remains {\em constant}
along the entire cascade. We will see
in \sref{section_wall} how the gauge theory for a D3-brane probing
more general geometries, such as a complex cone over the Zeroth
Hirzebruch surface, can exhibit a dramatically different behavior.
%

\section{Phases of $F_0$}

\label{section_F0}

We are now well-equipped with techniques of computing anomalous
dimensions, of duality trees and duality cascades.
Let us now initiate the study of some more complicated gauge theories.
Our first example will be the D-brane probe theory on a complex cone
over the zeroth Hirzebruch surface $F_0$, which is itself simply
$\IP^1 \times \IP^1$. This is a toric variety and the gauge theory was
analysed in \cite{toric}.

There are some reasons motivating the choice of this theory. The first
is its relative simplicity. The second is that its Seiberg dual phases
generically have multiplicities of
bifundamental fields greater than 2, whereby providing some
interesting properties. Indeed,
from the general analysis of \cite{fiol,HW},
a qualitative change in a RG flow towards the UV 
behavior is expected when such a multiplicity is exceeded. 
Finally, as we will discuss later, this theory admits
the addition of fractional branes. The presence of fractional branes
turns the theory non-conformal, driving 
a non-trivial RG flow. All together, this theory is a promising
candidate for a rich RG cascade structure.  

\begin{figure}[ht]
  \epsfxsize = 12cm
  \centerline{\epsfbox{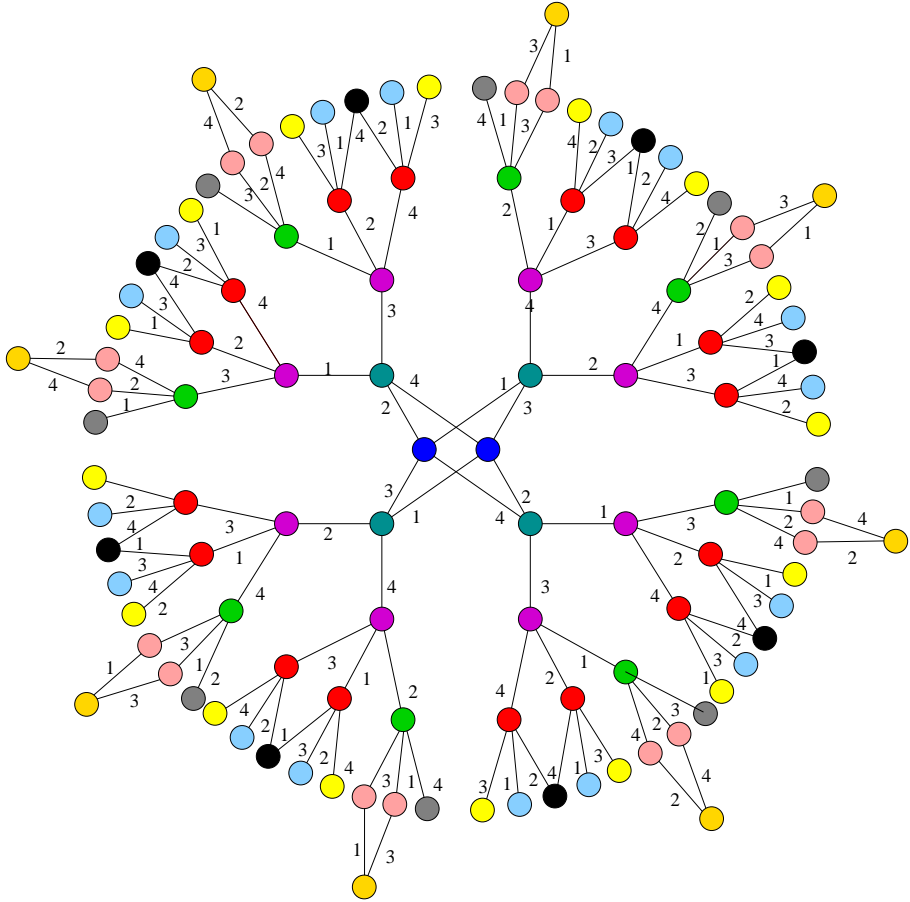}}
  \caption{The ``duality tree'' of Seiberg dual theories for $F_0$, it
  is in the shape of a ``flower,'' the {\it Flos
	Hirzebruchiensis}.}
  \label{F0_tree}
\end{figure}

\begin{figure}[ht]
  \epsfxsize = 10cm
  \centerline{\epsfbox{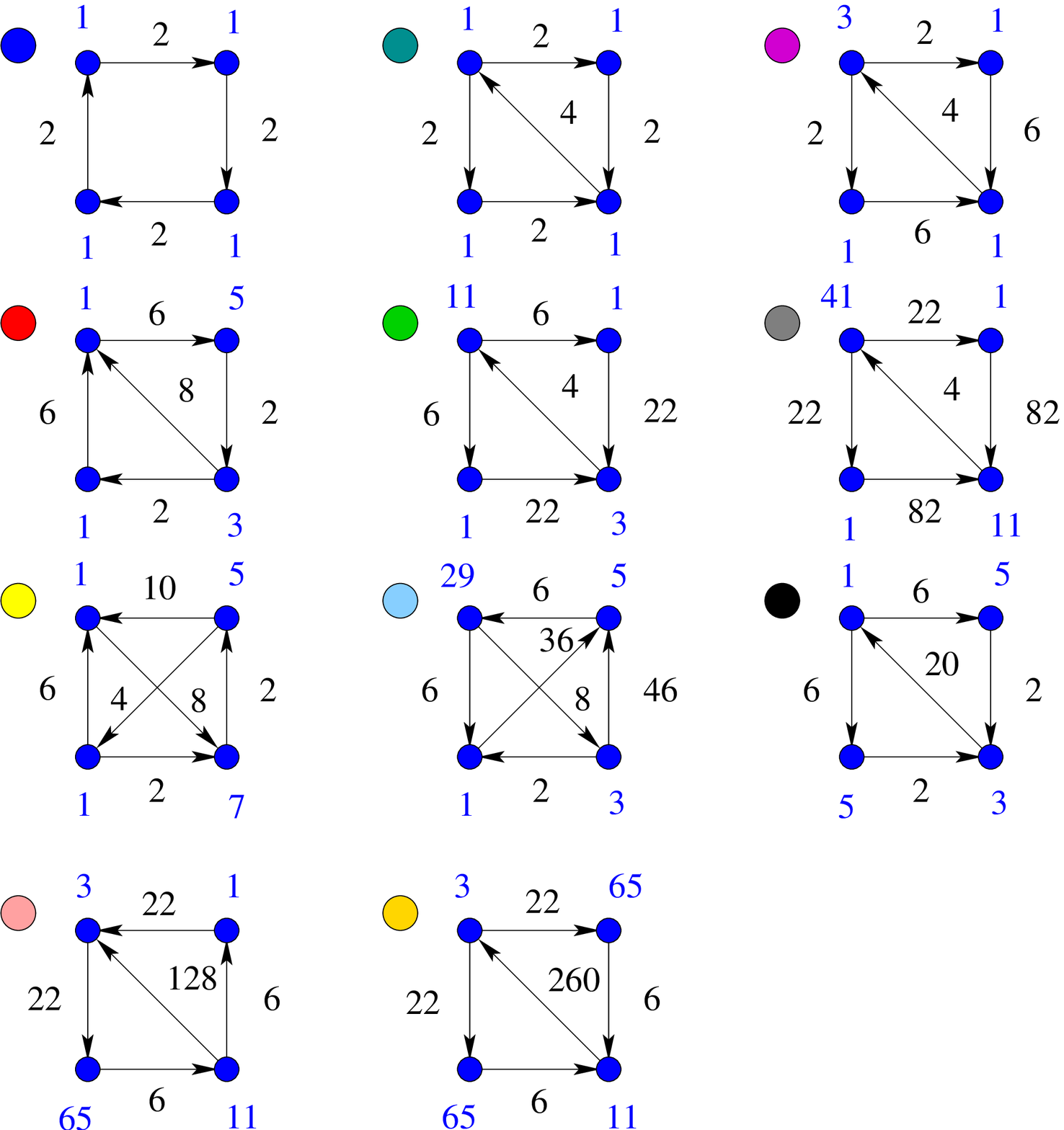}}
  \caption{Some Seiberg dual phases for $F_0$.}
  \label{rules_quivers_F0}
\end{figure}

The duality tree in this case is shown in \fref{F0_tree}; we shall
affectionately call it the ``$F_0$ flower,'' of the genus {\it Flos
Hirzebruchiensis} and family {\it Floris Dualitatis}.
We have drawn sites that
correspond to different theories with different colours; 
the colour-coded theories are summarized 
in \fref{rules_quivers_F0}.
Since the quiver has four gauge groups, there are four 
possible ways of performing Seiberg duality and thus there are four
branches coming out from each site of the tree. The numbers on each
branch corresponds to the node which was dualised.
A novel
point that was not present in the tree for $dP_0$ is the existence of
closed loops.

The possible existence of RG flows corresponding to these
closed loops is constrained by the requirement that the number
of degrees of freedom is increased towards the UV, in
accordance to the c-conjecture/theorem.
As it was stressed for the $dP_0$ and the conifold examples, the
duality tree for $F_0$ merely represents the infinite set 
of conformal theories which are Seiberg duals. 
Non-vanishing beta functions and the subsequent RG flow
are generated when fractional branes are included in the system. 
We reiterate this point: {\em the duality tree describes
duality cascades modulo fractional branes.} In other words, it
represents a ``projection" of actual cascades
to the space of vanishing fractional branes.
%

\subsection{$F_0$ RG flows}

In this section we will follow the RG flow towards the UV of the
theory living on D3-branes probing $F_0$, with the addition of 
fractional branes to obtain a non-conformal theory. 
As in the conifold example, the possible anomaly free probe and
fractional branes are determined by finding the integer 
null space of the intersection matrix that defines the quiver
\eref{anofree}. This can be 
done for any of the dual quivers that appear in the duality tree, but
the natural choice is the simplest of the $F_0$ quivers as was done in
\cite{fhhi} which is shown in \fref{rules_quivers_F0} as the first one
(blue dot).
The intersection matrix for this quiver is given by
\beq
A_{ij} =
\mat{0 & 2 & 0 &  0 \cr
0 & 0 &  2 &  0 \cr
0 & 0 & 0 &  2 \cr 
2 & 0 & 0 & 0} \Rightarrow
f_{ij} =\left( \begin{array}{cccc} 
0 & 2 & 0 &  -2 \\
-2 & 0 &  2 &  0 \\
0 & -2 & 0 &  2 \\ 
2 & 0 & -2 & 0 
\end{array} \right) \ .
\label{F0_intersection}
\eeq
A suitable basis for the nullspace of \eref{F0_intersection} is
$v_1=(1,1,1,1)$ and $v_2=(0,1,0,1)$. Therefore, the most generic 
ranks for the nodes in the quiver, consistent with anomaly
cancellation, are 
\beq
(n_1,n_2,n_3,n_4)=N(1,1,1,1)+M(0,1,0,1) \ .
\label{fractional_F0}
\eeq

Following the discussion in \sref{section_conifold_cascade},
we will refer to $N$ as the number of probe branes and $M$ as the
number of fractional branes. 
The theory is then conformal for $M=0$ and non-conformal otherwise. 
For $M \neq 0$, there will exist an RG cascade. 
The specific path to the UV is determined by the initial conditions of
the flow, namely the gauge couplings at  
a given scale $\Lambda_0$. As we will see, very different qualitative
behaviours can be obtained, depending on these initial conditions.

A crucial step in solving the conifold cascade was the identification
of a symmetry that provided us with control over the 
anomalous dimensions in the $M/N \ll 1$ limit. 
It is possible to find an analogous symmetry for $F_0$. Examining
\eref{fractional_F0} we see that this theory is invariant under
\begin{eqnarray}
M \rightarrow -M \\ \nonumber
N \rightarrow N+M \ .
\label{symmetry_F01}
\end{eqnarray}
In the limit $M/N  \ll 1$, this symmetry transformation becomes
\begin{eqnarray}
&&M \rightarrow -M \\ \nonumber
&&N \rightarrow N \ .
\label{symmetry_F02}
\end{eqnarray}

Up to now, we have only considered a single theory (the first quiver in
\fref{rules_quivers_F0}) and showed that for $M/N  \ll 1$ it is
symmetric under \eref{symmetry_F02}. But the whole tree of Seiberg dual
theories can be constructed using this model as
the starting point. Since the transformations in \eref{symmetry_F02} map
the initial theory onto itself, the models derived
from it by Seiberg duality are also invariant. Therefore,
\eref{symmetry_F02} constitute a symmetry
of the entire tree.

In analogy to the conifold case this symmetry implies that, for all the
dual theories, the odd order terms in the $M/N$
expansion of the anomalous dimensions of all their bifundamental fields
vanish and thus
\beq
\gamma=\gamma_c+O(M/N)^2 \ .
\eeq
This means that the leading, $O(M/N)$, non-zero contribution to the beta
functions can be computed using the anomalous
dimensions calculated at the conformal point. Schematically,
\beq
\beta(\gamma)=\beta(\gamma_c)+O(M/N)^2
\eeq
for all the beta functions.

The procedure outlined in \sref{section_computing_gammas} 
can thus be applied to determine the conformal anomalous dimensions,
which then can be used to work out the beta functions in the limit
$M/N \ll 1$ and study the running of the gauge couplings
as we flow to the UV. Let us do so in detail.
The beta-functions in \eref{betas}, for a quiver theory with $k$ gauge
group factors, ranks $\{n\}_i$, adjacency matrix $A_{ij}$ and loops
indexed by $h$ corresponding to gauge invariant operators that appear
in the superpotential, now becomes
\bea
\beta_{i \in \mbox{nodes}}
&=& 3 n_i -\frac12 \sum_{j=1}^k (A_{ij} + A_{ji}) n_j + \frac12
	\sum_{j=1}^k (A_{ij} \gamma_{ij} + A_{ji} \gamma_{ji}) n_j \nn \\
\beta_{h \in \mbox{loops}} 
&=& -d(h) + \frac12 \sum_{h} \gamma_{h_i h_j} \ ,
\label{betasquiver}
\eea
where in the second expression $\beta_{h \in \mbox{loops}}$ associated
with the terms in the superpotential the index in the 
sum over $\gamma_{h_i h_j}$ means consecutive arrows in a loop and
$d(h)$ is determined by 3 minus the number of fields in the loop.

We will make liberal use of \eref{betasquiver} throughout. For our
example for the first phase of $F_0$, 
the ranks $(n_1, n_2, n_3, n_4) = (1,1,1,1)$, together
with intersection matrix from \eref{F0_intersection},
\eref{betasquiver} reads
\beq
\label{betasF0}
1 + \gamma_{1,2} + \gamma_{4,1} = 0, \quad
1 + \gamma_{1,2} + \gamma_{2,3} = 0, \quad
1 + \gamma_{2,3} + \gamma_{3,4} = 0, \quad
1 + \gamma_{3,4} + \gamma_{4,1} = 0, \quad
1 + \frac12 (\gamma_{1,2} + 
	\gamma_{2,3} + \gamma_{3,4} + \gamma_{4,1}) = 0 \ ,
\eeq
which affords the solution
\beq
\label{solF0-1}
\{
 \gamma_{1,2} \rightarrow -1 - \gamma_{4,1}, \quad
 \gamma_{2,3} \rightarrow \gamma_{4,1}, \quad
 \gamma_{3,4} \rightarrow -1 - \gamma_{4,1}
\} \ .
\eeq
We see that there is one undetermined $\gamma$. To fix this we appeal
to the maximisation principle presented in
\sref{section_computing_gammas}. The central charge
\eref{definition_a_2} now takes the form (where $\gamma_{4,5}$ is
understood to mean $\gamma_{4,1}$.
\beq
\label{aF0-1}
a = \frac{3}{4} + \frac{1}{16} 
\sum_{i=1}^4
\left( 1 + 
\frac{{\left( -1 + \gamma_{i,i+1} \right)}^3}{3} 
- \gamma_{i,i+1} \right) \ .
\eeq
Upon substituting \eref{solF0-1} into \eref{aF0-1}, we obtain
\beq
a(\gamma_{4,1}) = -\frac38 (-2 + \gamma_{4,1} + \gamma_{4,1}^2) \ ,
\eeq
the maximum of which occurs at $\gamma_{4,1} = -\frac12$.
And so we have in all, upon using \eref{solF0-1},
\beq
\label{solF0-2}
\gamma_{1,2}=\gamma_{2,3}=\gamma_{3,4}=\gamma_{4,1}=-1/2 \ .
\eeq

Let us remark, before closing this section, that there is
an alternative, though perhaps less systematic, 
procedure to determine anomalous  
dimensions that does not rely on the maximization of $a$. 
For every theory in the $F_0$ cascade the 
space of solutions to \eref{betasF0} is one dimensional. Fixing this
freedom at 
any given point determines the anomalous dimensions 
in the entire duality tree. Maximization of the central charge $a$ is
a possible way of determining this free parameter. For $F_0$, a simple
alternative is to make use of the symmetries of
the theory (quiver and superpotential). 
Our theory \eref{F0_intersection} for example, instantly has all
$\gamma$'s equal by the $\IZ_4$ symmetry of the quiver. Therefore, in
conjunction with the solutions \eref{solF0-1} to conformality, gives 
\eref{solF0-2} as desired.
Once the anomalous dimensions of theory \eref{F0_intersection} 
are determined, the freedom
that existed in the conformal solutions of all the dual theories is fixed.
This is done by matching the scaling dimensions of composite Seiberg
mesons every time a Seiberg duality is performed and/or by noting that the
anomalous dimensions of fields that are neutral under the dualized gauge
group are unchanged.

\subsection{Closed Cycles in the Tree and Cascades}
\label{section_closed_cycle}
\begin{figure}[ht]
\[
\begin{array}{ccc}
{\epsfxsize=3cm\epsfbox{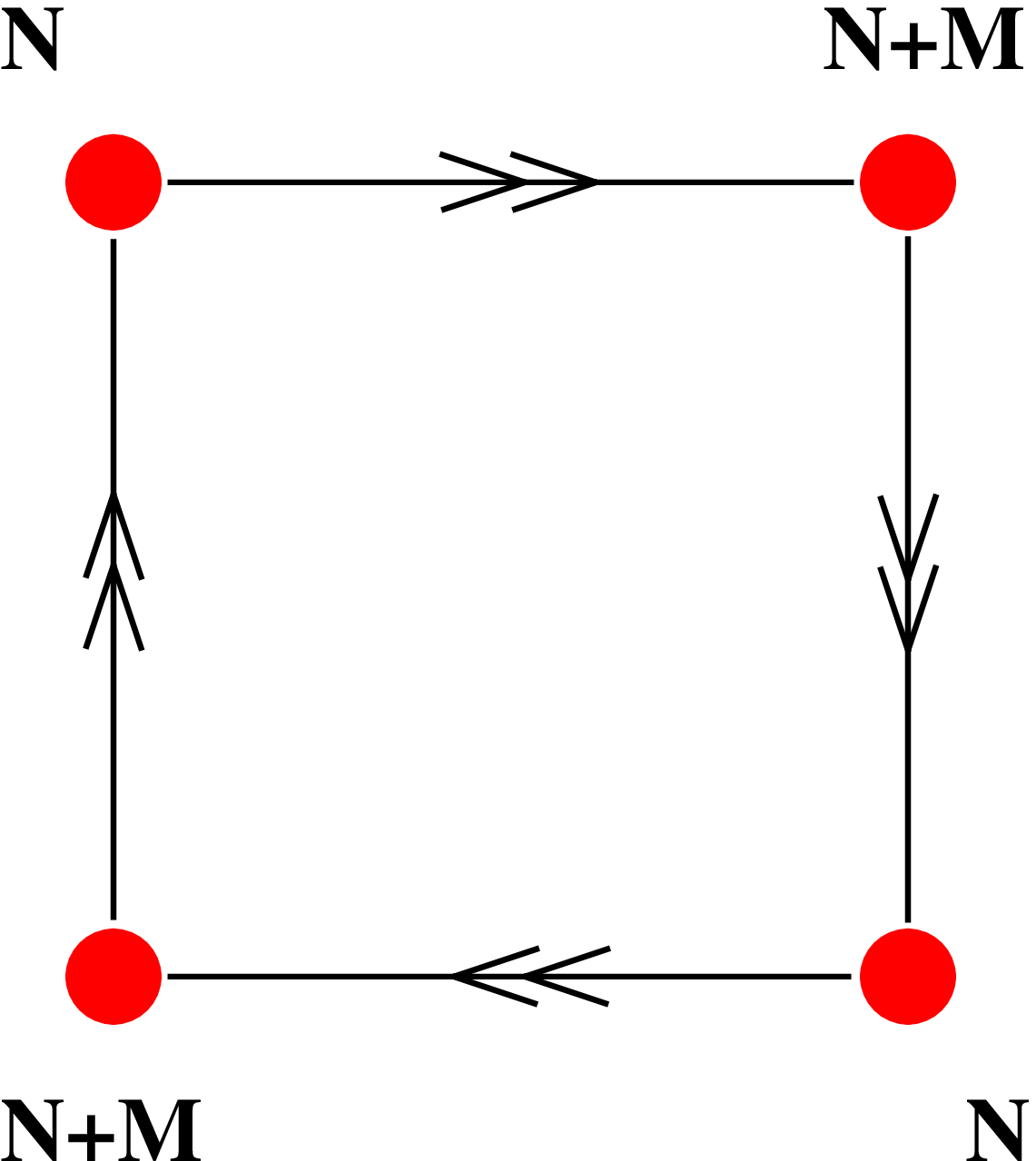}} 
& \qquad
& {\epsfxsize=3cm\epsfbox{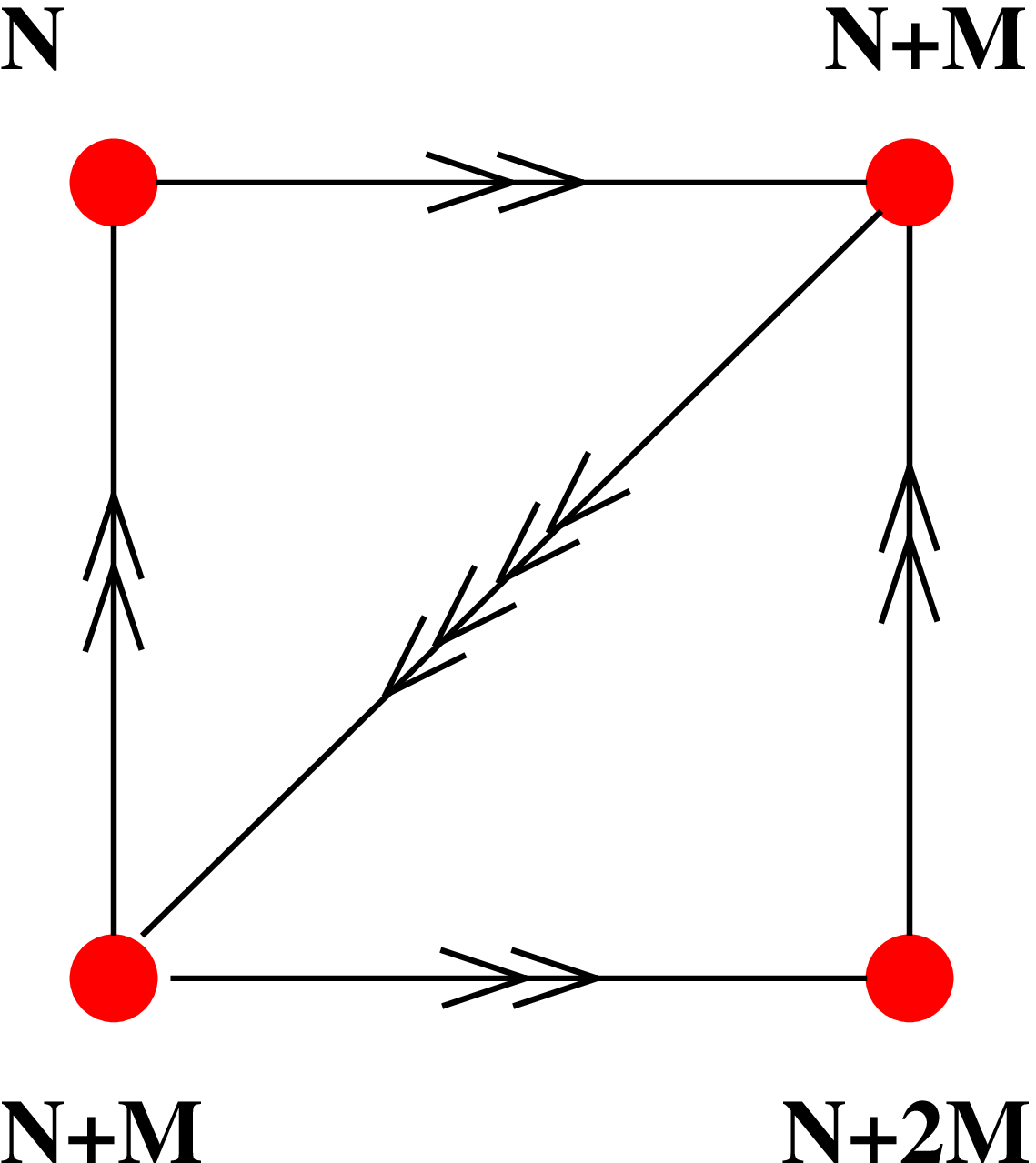}}  \\
\mbox{Model A} 
& \qquad
&
\mbox{Model B} 
\end{array}
\]
\caption{Quivers for Models A and B. Model A corresponds to the
	choice of ranks
	$(n_1,n_2,n_3,n_4)_A=N(1,1,1,1)+M(0,1,0,1)$, from which
	model B is obtained by dualizing node 3. It has ranks
	$(n_1,n_2,n_3,n_4)_B=N(1,1,1,1)+M(0,1,2,1)$ \ .
\label{toric_quivers_F0}
}
\end{figure}

Now we wish to find the analogue of the conifold cascade in
\sref{section_conifold_cascade} here. For this we wish to look for
``closed cycles'' in the duality tree (\fref{F0_tree}). In the
conifold case the theory was self dual and we cascaded by adding
appropriate fractional branes. Here we do indeed see a simple cycle
involving 2 sites (namely the blue and the dark green). We will call
these two theories models $A$ and $B$ respectively and draw them in 
\fref{toric_quivers_F0}. Model $A$ is the example we addressed above.

The starting point 
will be model $A$, its superpotential and the set of gauge
couplings at a scale $\Lambda_0$.
We recall from \eref{solF0-2} that the anomalous dimensions at the 
conformal point are
$\gamma_{1,2}=\gamma_{2,3}=\gamma_{3,4}=\gamma_{4,1}=-1/2$. 
This leads to the following values for the beta functions for the 4
gauge group factors:
\beq
\begin{array}{lcl}
\beta_1=-3M & \ \ \ \ \ & \beta_2=3M \\
\beta_3=-3M & \ \ \ \ \ & \beta_4=3M \ .
\end{array}
\label{beta_toricF0_1}
\eeq

These beta-functions are constants, which means that
the running of $x_i$, the inverse squared
couplings as a function of the log scale 
is linear, with slopes given by \eref{beta_toricF0_1}. Let us thus
run $x_i$ to the UV accordingly.
We see that $\beta_1$ and $\beta_3$ are negative so at
some point the inverse couplings for the first or the third node will
reach 0. Which of them does so first depends
on the value of the initial inverse couplings
we choose for $n_1$ and $n_3$. We dualise the
node for which the inverse coupling first reaches 0, say node 3.
This will give us Model $B$. If instead node 1 has the inverse coupling
going to 0 first, we would dualise on 1 and obtain
a theory that is equivalent to Model $B$ after a reflection of the quiver
(we can see this from \fref{F0_tree}).

Next we compute the anomalous dimensions for Model B at the
conformal point.  In analogy to \eref{betasF0} and \eref{aF0-1} we
now obtain 
$\gamma_{1,2}=\gamma_{3,2}=\gamma_{4,3}=\gamma_{4,1}=-1/2$ and
$\gamma_{2,4}=1$, which gives the beta functions for the next step:
\beq
\begin{array}{lcl}
\beta_1=-3M & \ \ \ \ \ & \beta_2=0 \\
\beta_3=3M & \ \ \ \ \ & \beta_4=0 \ .
\end{array}
\label{beta_toricF0_2}
\eeq 
From these we run the couplings at this stage again, find the node for
which the inverse coupling first goes to 0. And dualise that node.
We see a remarkable feature in \eref{beta_toricF0_2}.
To the level of approximation 
that we are using, only the first gauge group factor has a negative beta
function. This implies that the next node to be dualized is precisely
node 1. Performing Seiberg duality thereupon takes us
to a quiver that is exactly of the form of Model A,
only with the ranks differing in contributions proportional to 
$M$, i.e., different fractional brane charges

\begin{figure}[ht]
  \epsfxsize = 10cm
  \centerline{\epsfbox{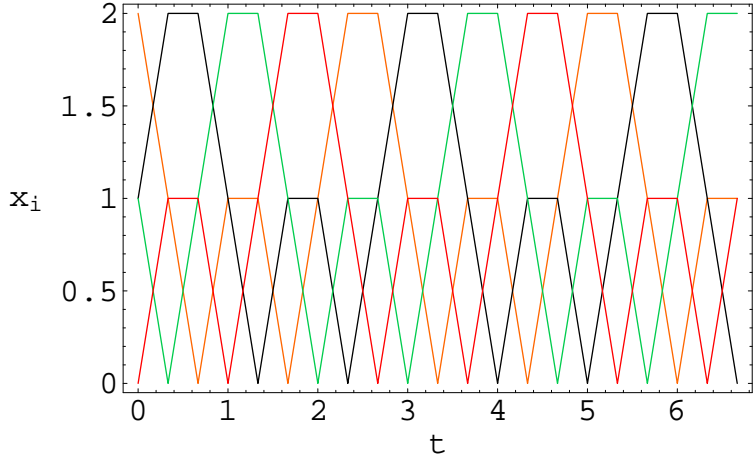}}
  \caption{Duality cascade alternating between the A and B toric
  models. 
The colouring scheme is such that orange, black, green, and red
respectively represent nodes 1, 2, 3 and 4.
}
  \label{F0Toric_cascade}
\end{figure}

By iterating this procedure it is possible to see that the entire
cascade corresponds to a chain that alternates 
between type A and type B models. Furthermore, the length of the even
steps of the cascade,  measured on the $t = \log \mu$ axis is constant.
The same statement applies to the length of the odd steps.
This cascade is presented in \fref{F0Toric_cascade} for the initial
conditions $(x_1,x_2,x_3,x_4)=(2,1,1,0)$.
%

\subsection{Duality wall}

\label{section_wall}

\begin{figure}[ht]
  \epsfxsize = 3cm
  \centerline{\epsfbox{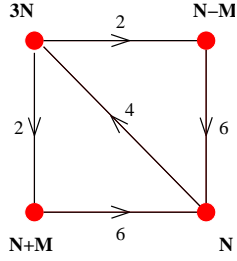}}
  \caption{Model $C$ for the $F_0$ theory. It is obtained from dualising
  node 2 and then 1 from the simplest Model $A$.}
  \label{quiver_F0_3}
\end{figure}

We have seen in \sref{section_closed_cycle} that models A and B form a
closed cascade and are not connected 
to the other theories in the $F_0$ duality tree by the RG flow,
regardless of initial conditions. This motivates
the study of duality cascades having other Seiberg dual theories as
their starting points. The simplest choice 
corresponds to the model in \fref{quiver_F0_3}. This theory is
obtained from Model $A$ by Seiberg dualizing node 2 followed by  1. We
will call this Model $C$.
\subsubsection{Decreases in the Step}
Applying the formalism developed in previous sections we can proceed
to compute, for any initial condition, the RG cascade  
as the theory evolves to the UV. 
The starting point is the computation of the anomalous dimensions for
Model $C$ at the conformal point. These, by the techniques above, 
turn out to be $\gamma_{1,2}=\gamma_{1,4}=-3/2$,
$\gamma_{2,3}=\gamma_{4,3}=5/2$ and 
$\gamma_{3,1}=-1$. Using them to calculate the beta functions, we
obtain
\beq
\begin{array}{lcl}
\beta_1=0 & \ \ \ \ \ & \beta_2=-3M \\
\beta_3=0 & \ \ \ \ \ & \beta_4=3M \ .
\end{array}
\label{beta_F0_3}
\eeq

\begin{figure}[ht]
\begin{center}
\[
\begin{array}{cccc} 
\mbox{Couplings vs. step} 
&
\qquad
&
\mbox{Couplings vs. scale}
\\
{\epsfxsize=8cm\epsfbox{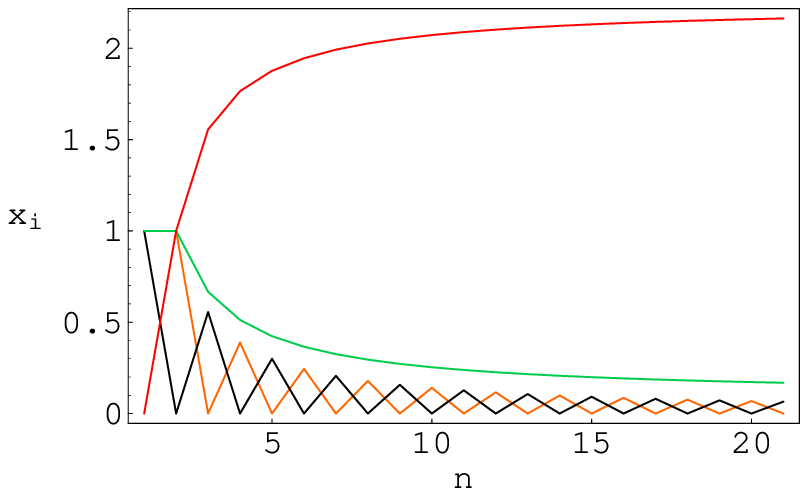}} &
\qquad
&
{\epsfxsize=8cm\epsfbox{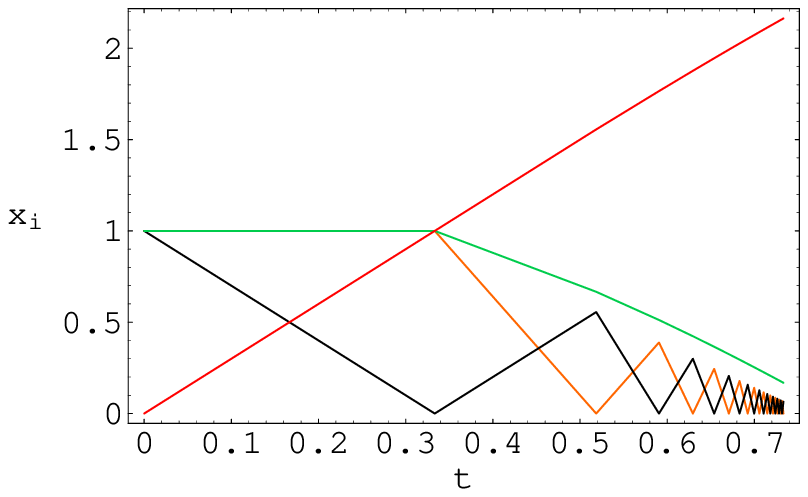}}  
\end{array}
\]
\end{center}
\caption{Evolution of gauge couplings with {\bf (a)} the step
in the duality cascade and {\bf (b)} the energy scale for initial
conditions $(x_1,x_2,x_3,x_4)=(1,1,1,0)$. 
The colouring scheme is such that orange, black, green, and red
respectively represent nodes 1, 2, 3 and 4.
\label{F0_cascade_1}}
\end{figure}

\begin{figure}[ht]
\begin{center}
\[
\begin{array}{ccc} \mbox{$\Delta$ vs. step} & 
\qquad
& \mbox{Scale vs. step}
\\
{\epsfxsize=8cm\epsfbox{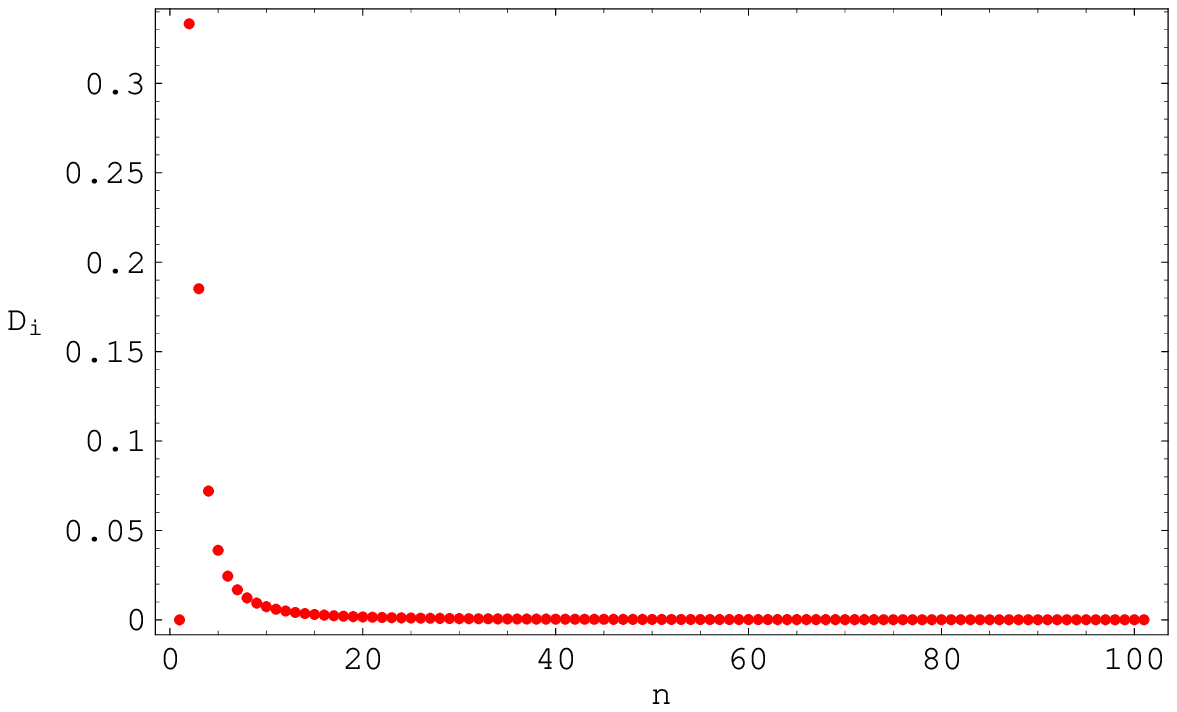}} &
& \qquad
{\epsfxsize=8cm\epsfbox{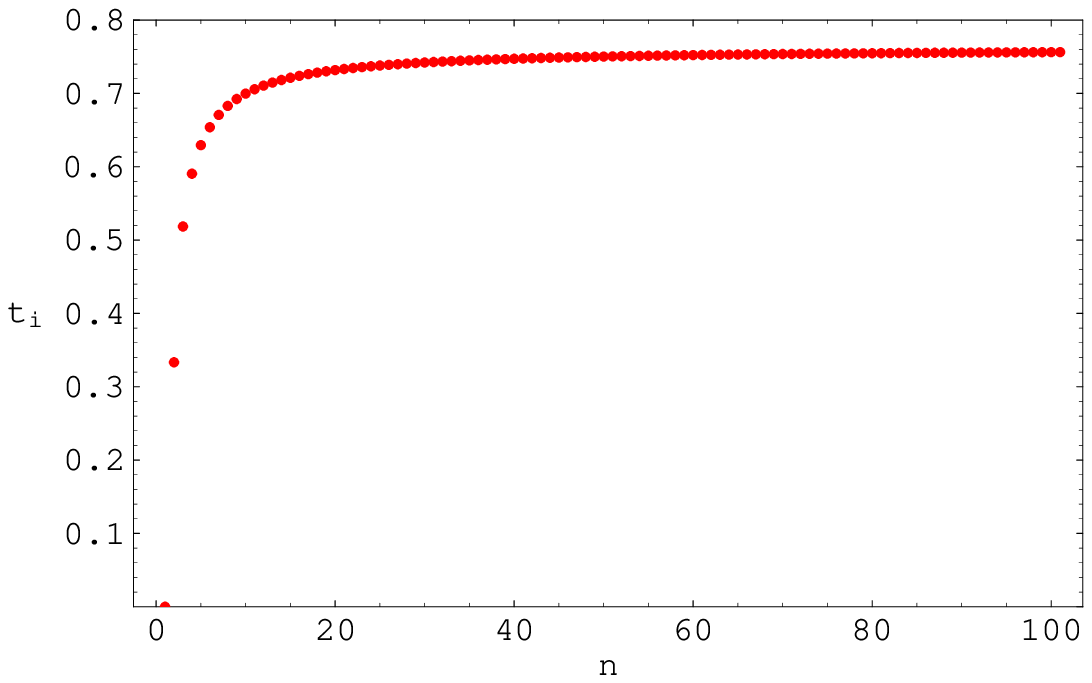}}  
\end{array}
\]
\end{center}
\caption{The evolution of $\Delta$, the size of the increment during
each dualisation and the energy scale increase as we dualise, for the
initial conditions
$(x_1,x_2,x_3,x_4)=(1,1,1,0)$.
\label{del-scale_F0_1}
}
\end{figure}

With these let us evolve to the UV.
Let us first consider the case in
which the initial condition for the inverse gauge couplings are 
$(x_1,x_2,x_3,x_4)=(1,1,1,0)$. 
\fref{F0_cascade_1}  
shows the evolution of the four inverse gauge couplings both as a
function of the step in the cascade and as a function  
of the logarithm of the scale. 
An interesting feature is that the distance, $\Delta_i$, 
between successive dualizations is monotonically decreasing. This 
marks a departure from the behaviors observed in the conifold cascade
and from the example presented in \sref{section_closed_cycle},
where $\Delta_i$ remained constant (cf.~\fref{couplings_conifold}). 
However, this fact does not necessarily mean the 
convergence of the dualization scales. Indeed, we
plot the intervals $\Delta_i$ in \fref{del-scale_F0_1}.a while 
\fref{del-scale_F0_1}.b shows the resulting dualization scales. 
The slope of this 
curve is decreasing, reflecting the decreasing behavior of
$\Delta_i$. 
Nevertheless, 
{\it ad infinitum}, the scale may diverge.
%
%
\subsubsection{A Duality Wall}
\begin{figure}[ht]
\begin{center}
\[
\begin{array}{ccc} 
\mbox{Couplings vs. step} & 
\qquad
& \mbox{Couplings vs. scale}
\\
{\epsfxsize=8cm\epsfbox{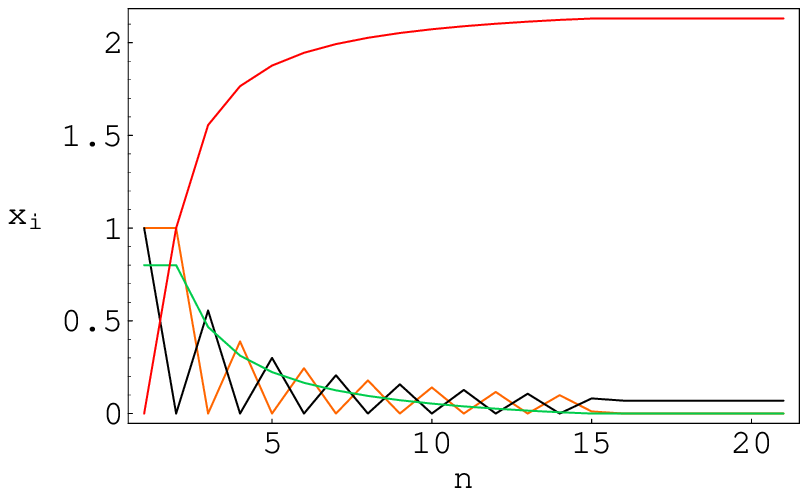}} &
\qquad
&
{\epsfxsize=8cm\epsfbox{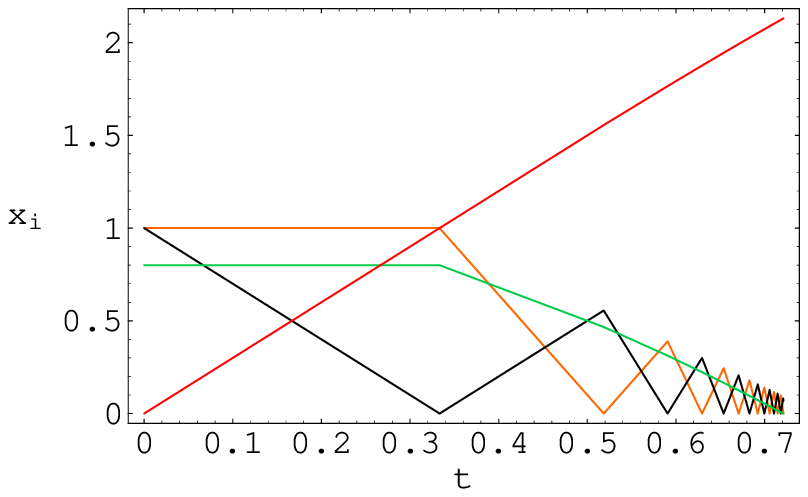}}  
\end{array}
\]
\end{center}
\caption{Evolution of gauge couplings with {\bf (a)} the step
in the duality cascade and {\bf (b)} the energy scale for initial
conditions $(x_1,x_2,x_3,x_4)=(1,1,4/5,0)$.
The colouring scheme is such that orange, black, green, and red
respectively represent nodes 1, 2, 3 and 4.
\label{F0_cascade_2}
}
\end{figure}

Let us now consider a different set of initial conditions, given by
$(x_1,x_2,x_3,x_4)=(1,1,4/5,0)$. The flow of the 
inverse couplings is now shown in \fref{F0_cascade_2}. 

A completely new phenomenon appears in this case. 
Something very drastic happens after the 14-th step in the
cascade.  As a consequence of lowering the initial value of $x_3$, 
the third node gets dualized at this step, producing an explosive 
growth of the number of chiral and vector multiplets in the quiver. 
This statement can be made precise: {\em at the 14-th step 
node 3 is dualized and the subsequent quivers have all their
intersection numbers greater than 2}.
In this situation the results of \cite{fiol,HW} suggest that
a duality wall is expected. This phenomenon is characterized by
a flow of the dualization scales towards an UV {\em accumulation
point} with an exponential divergence in the number of degrees of
freedom. This 
asymptotic behavior can be inferred once Seiberg dualities are interpreted
as Picard-Lefschetz monodromy transformations which we discuss in  
Appendix 2 (q.v.~\eref{accupoint}).

\fref{F0_cascade_2} shows a very small running of the gauge couplings
beyond this point. 
This is not due to a vanishing of the beta functions, 
but to the extreme reduction of the length of 
the $\Delta_i$ intervals.

%

%
\begin{figure}[ht]
\begin{center}
\[
\begin{array}{ccc} 
\mbox{$\Delta$ vs. step} &
\qquad
& \mbox{Scale vs. step}
\\
{\epsfxsize=8cm\epsfbox{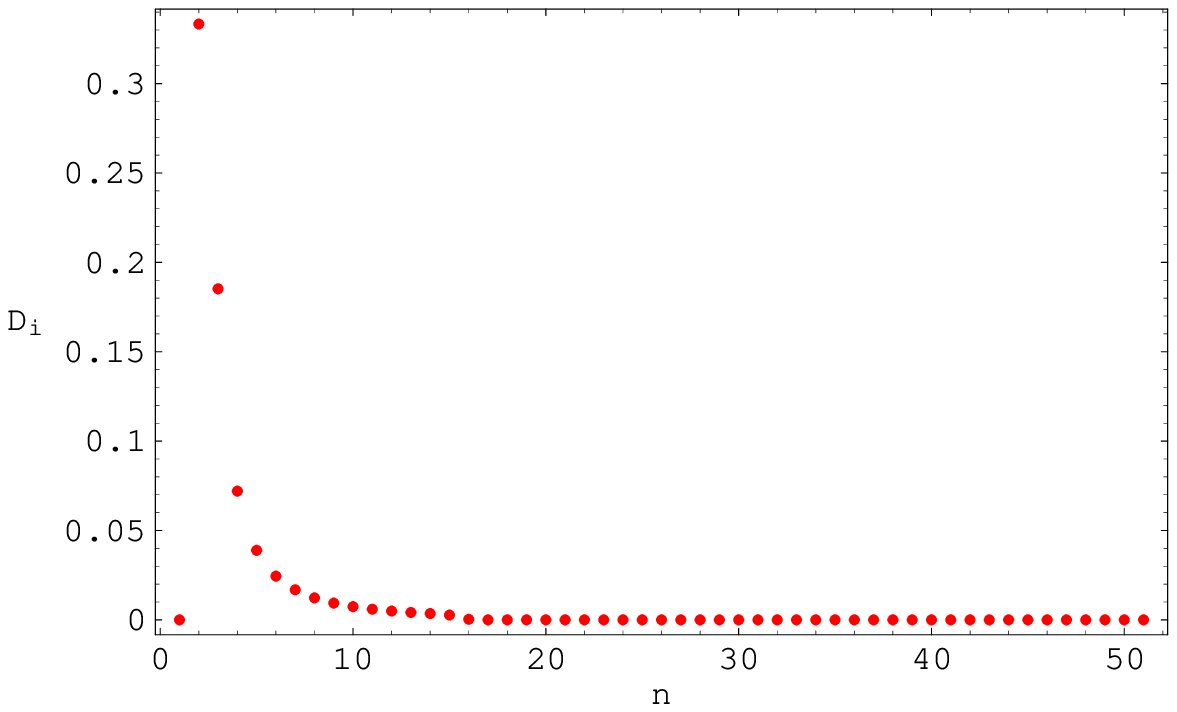}} &
\qquad
&
{\epsfxsize=8cm\epsfbox{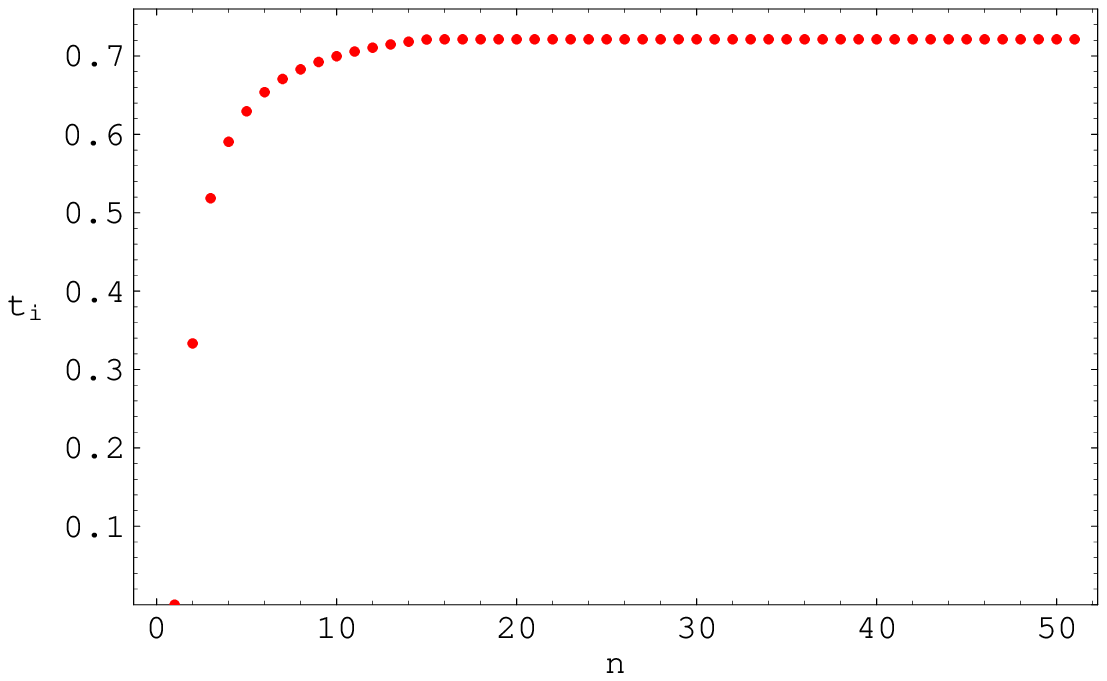}}  
\end{array}
\]
\end{center}
\caption{The evolution of $\Delta$, the size of the increment during
each dualisation and the energy scale increase as we dualise, for the
initial conditions
$(x_1,x_2,x_3,x_4)=(1,1,4/5,0)$.
\label{del-scale_F0_2}
}
\end{figure}
In contrast to \fref{del-scale_F0_1}, for the initial conditions
$(1,1,4/5,0)$, we have drawn the plots in \fref{del-scale_F0_2}.a and
\fref{del-scale_F0_2}.b. Both of them indicate that a 
limiting scale which cannot be surpassed is 
reached as the theory flows towards the UV. This is precisely what we
call a {\bf duality wall.}

%
%
\subsection{Location of the Wall}
We have just seen that starting from the quiver in
\fref{quiver_F0_3} for $F_0$ 
with initial conditions $(x_1,x_2,x_3,x_4)=(1,1,4/5,0)$ 
a duality wall is reached.
Let us briefly examine
the sensitivity of the location of the duality wall 
to the initial inverse couplings.

\begin{figure}[ht]
  \epsfxsize = 19cm
  \centerline{\epsfbox{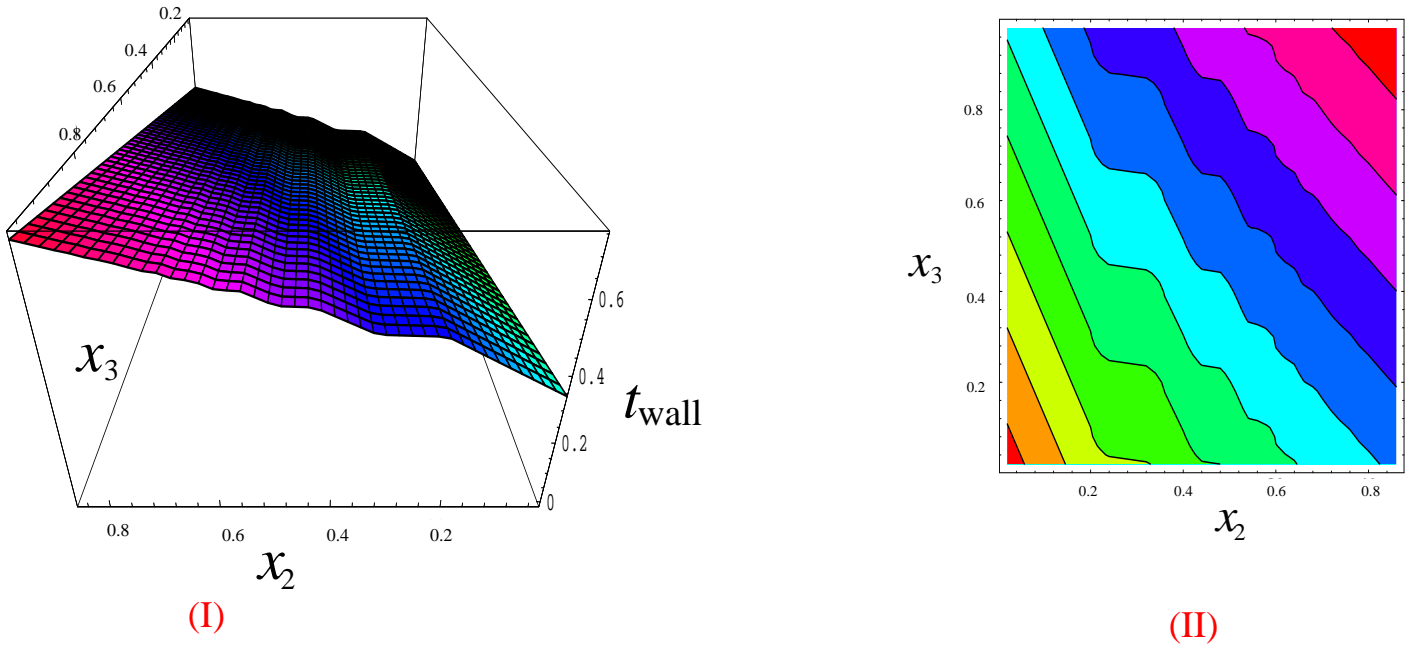}}
  \caption{A plot of the position $t_{wall}$ against the initial gauge coupling
  values $(1, x_2, x_3, 0)$. (I) is the 3-dimensional plot and (II) is
  the contour plot versus $x_2$ and $x_3$.}
  \label{f:pos}
\end{figure}

Let our initial inverse gauge couplings be $(1, x_2, x_3, 0)$, with $0
< x_2,x_3 < 1$, and we repeat the analyses in the previous two
subsection. We study the running of the beta functions, and determine
the position of the duality wall, $t_{wall}$, for various initial values.
We plot in \fref{f:pos}, the position of the duality wall against
the initial values $x_2$ and $x_3$, both as a three-dimensional plot
in I and as a contour plot in II.
We see that the position is a step-wise function. 
A similar behavior has been already observed 
in \cite{HW} for $dP_0$ in the case of vanishing anomalous dimensions.
%
%
%

%
\subsection{A $\IZ_2$ Symmetry as T-Duality}
A remarkable symmetry seems to be captured by the gauge
theory discussed above. Suppose that,
starting from Model $C$, we had decided to follow the RG flow towards
the IR instead of the UV. There is a simple trick that
accomplishes this task, namely to look at the flow to the
the UV of a theory in which the beta functions have changed sign and
where the log-scale $t$ 
has been replaced by $-t$. Since we are considering beta functions that
are
linear in the number of fractional branes $M$, changing the sign of the
beta functions can be interpreted as changing $M$ to $-M$.
However, upon inspecting \fref{quiver_F0_3}, 
we see that $M \leftrightarrow -M$ is
nothing more than a $\IZ_2$ reflection of the quiver along the $(13)$
axis.
Therefore, the flow to the IR that starts from Model $C$ is simply the
flow to the UV of its $\IZ_2$ reflection. Therefore, the whole cascade
we have already computed also describes, upto this reflection, the
cascade to the IR. 

Let us elaborate on the origin of this $\IZ_2$ symmetry. In the
holographic
dual of the gauge theory, the energy scale $\mu$ is typically associated
to a radial coordinate $R$. Then, transforming $t=\log \mu$ to $-t$
corresponds to an inversion of the radial coordinate $R \leftrightarrow
1/R$ in the holographic dual theory. Therefore this $\IZ_2$ symmetry
displayed by the 
gauge theory RG flow indicates a $\IZ_2$ T-duality-like
symmetry of string theory on the underlying geometry.
The scale of model $C$ can then be interpreted as the self dual radius
of the dual holographic theory. This holographic theory then exhibits a
minimal length beyond which there are no further new phenomena. Every
physical quantity at a scale less than this scale can be expressed in
terms of a different physical quantity by applying the $\IZ_2$ action
described above. 
It will be very interesting to explore this symmetry further.

\section{Phases of $dP_1$}
\label{sec:dP1}
We have initiated the study of duality walls for general quiver
theories and in the foregoing discussion analysed in detail the
illustrative example of the cone over the zeroth Hirzebruch surface.
It is interesting to extend the construction of duality trees to other
gauge theories and the structures and RG flows that may emerge. 

\begin{figure}[ht]
  \epsfxsize = 10cm
  \centerline{\epsfbox{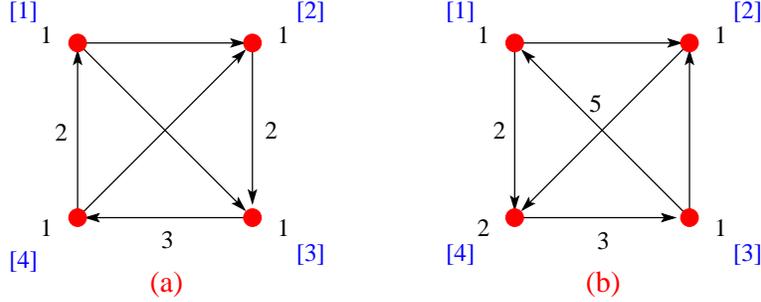}}
  \caption{The quivers for the gauge theory arising from $dP_1$. (a)
  is the toric phase while (b) is obtained from (a) by dualising
  either node 3 or 4 and is a non-toric phase. The ranks of the
  nodes are
  denoted by blue square brackets.}
  \label{quiver_dP1}
\end{figure}

\begin{figure}[ht]
  \epsfxsize = 9cm
  \centerline{\epsfbox{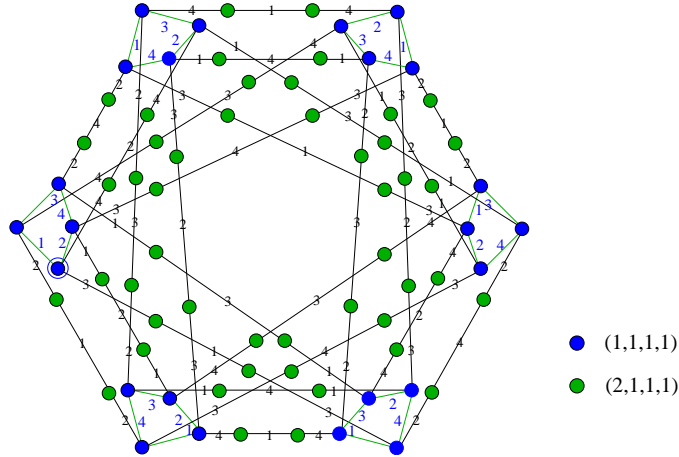}}
  \caption{The tree of Seiberg dual theories for $dP_1$. There are six
  toric islands (sets of blue sites) in this tree.}
  \label{dP1_tree}
\end{figure}

Let us briefly consider perhaps the next simplest case,
viz., the gauge theory on a D3-brane probing a complex cone
over $dP_1$, the first del Pezzo surface, which is $\IP^2$ blown up at
1 point.
This is again a toric variety and the
theory has been extensively studied \cite{toric,symmetries}. 
There is only one toric phase in this case\footnote{We follow here the
	nomenclature  
	of \cite{symmetries}, where a theory was denoted toric if it
	had all its gauge group factors equal to $U(N)$, i.e., all the
	ranks are equal.} ,  
whose quiver is shown in \fref{quiver_dP1}; the ranks are $(1,1,1,1)$.
This model is self-dual under the dualizations of nodes 1 or 2,
and is transformed into a theory with gauge group $U(2N) \times U(N)
\times U(N) \times U(N)$ when dualizing any of the
other two nodes. Again using the notation of \eref{fractional_F0}, we
denote the ranks in the conformal case, as
$(2,1,1,1)$.
The duality tree obtained by taking into account these two models is
shown in \fref{dP1_tree}.
The encircled theory is the one in \fref{quiver_dP1}.a and each link
shows the
associated dualized node. Other  
Seiberg dualizations of the $(2,1,1,1)$ model lead to $(5,2,1,1)$ and
$(4,2,1,1)$ theories.
We have not included them in \fref{dP1_tree} for simplicity.
Already at this level we can see that the duality space of $dP_1$ is
very rich, exhibiting a phenomenon that 
we will call {\bf ``toric islands''} (isolated sets of toric models). 
Furthermore, the abundance of closed loops makes it quite 
different from the $dP_0$ and $F_0$ cases. The appearance of a
six-fold multiplicity of islands is a result 
of the combinatorics that gives the possible reorderings of a given
quiver. Explicitly, the number of possible re-orderings 
of the four nodes in the toric quiver are $4!=24$. After grouping
these toric models in islands of four, we are lead 
to the six expected islands. 

\section{Conclusions and Prospects}
\label{sec:con}
In applying some recent technology to the classic conifold cascade
phenomenon \cite{klst}, 
we have here studied how duality walls arise in string
theory. Endowed with the systematic methods from toric duality
\cite{toric,symmetries} and maximisation principles in determining
anomalous dimensions in arbitrary four-dimensional SCFT's
\cite{Intri1,Intri2}, we have re-examined the ideas of \cite{fiol,HW}
by supplanting complete anomalous dimensions to the beta functions,
emerging under a fully string theoretic realisation.

Indeed, armed
with the vast database of ``duality trees'' encoding various
(generalised) Seiberg dual phases of four-dimensional 
SCFT ${\cal N}=1$ gauge theories that live on world-volumes of D3
branes probing various singular geometries, we have outlined a general
methodology of analysing cascading phenomena and finding duality
walls. At liberty to take advantage of the well-studied toric
singularties, we have 
used the enlightening example of the theory corresponding to the
complex cone $F_0$ over the zeroth Hirzebruch surface. We have shewn
how one adds fractional branes to take the theory out of conformality
and hence obtain a nontrivial running of the (inverse) gauge
couplings. Thereafter, by either symmetry arguments \cite{symmetries}
or maximisation of central charge $a$ \cite{Intri1}, one could determine
the exact form of the NSVZ beta-functions, in the limit where the
number of physical branes is much larger than that of fractional ones.
We have then shewn how to apply this running to consecutive
applications of Seiberg duality. 

We find, when one identifies
appropriate ``closed cycles'' in the
tree, generalisation of the conifold cascades. We also find
the existence of 
``duality walls'' (cf.~\fref{F0_cascade_2}).
This occurs whenever subsequent
applications of the duality (cascade) result in the rapid decrease in
the evolution, towards the UV, 
of the full beta functions, whereby creating an
accumulation point onto which all asymptotic values of the couplings
pile. Such a wall signifies a finite energy scale beyond which Seiberg
dualities cannot proceed and at which an infinite number of degrees of
freedom blossom in the string theory.

An interesting appearance is played by the $\IZ_2$ symmetry which the
beta function equations exhibit and the non-trivial action on the
scale. This action simply inverts the scale and defines a critical
scale, say 1, at which there is a reflection point in the physics. All
phenomena above this scale are related to those below it. In the dual
geometry this is reminiscent of the familiar T-duality and the
critical scale corresponds to the self-dual radius of T-duality. 
The position of the wall is sensitive to the initial
conditions, viz., the values of the gauge couplings, and exhibits a
step-wise behaviour with respect to them (cf.~\fref{del-scale_F0_2}).

Of course, our example is but the simplest of a wide class of
theories. We have briefly touched upon a more involved case of the
cone over the first del Pezzo surface. There, interesting ``toric
islands'' appear, giving an even richer structure for the cascade.
Indeed, a plethora of theories awaits our exploration.

As an aside, we have used the AdS/CFT diationary, which dictates in
particular that the central charge $a$ is inversely proportional to
the horizon volume in the dual bulk theory, to
obtain a formula applicable to all the cones over del Pezzo surfaces
(cf.~\eref{VdPn}).

\vspace{1in}

\section*{Acknowledgements}
We would like to thank Francis Lam for many enlightening
discussions and also U.~Gursoy,
C.~Nu\~{n}ez, M.~Schvellinger, and J.~Walcher for insightful
comments.
We sincerely acknowledge the gracious patronage of
the CTP and the LNS of MIT as well as
the department of Physics at UPenn. Further support is
granted from the U.S. Department of Energy under cooperative
agreements $\#$DE-FC02-94ER40818 and $\#$DE-FG02-95ER40893.
A.~H.~is also indebted to the Reed Fund Award
and a DOE OJI award, and Y.-H.~H., also to an NSF Focused Research
Grant DMS0139799 for ``The Geometry of Superstrings.''

\newpage
\section{Appendix 1: Picard-Lefschetz Monodromy Transformations}
In this Appendix we briefly remind the reader of Seiberg and toric
dualities from the perspective of Picard-Lefschetz monodromy and
$(p,q)$ seven-branes perspective. Now to engineer quiver theories of
gauge group $\prod\limits_{i=1}^n U(N_i)$ we can take a set of $n$
$(p,q)$
sevenbranes with charges $(p_i,q_i)$ and multiplicity $N_i$, 
corresponding to a
cycle $S_i$ in the singular geometry, satisfying
\beq
\sum_i^n N_i (p_i,q_i)=0 \ .
\label{zero_sum}
\eeq
The intersection matrix for the quiver is given by
\beq
I_{ij} = S_i \cdot S_j = \det \mat{p_i & q_i \cr p_j & q_j} \ .
\eeq
Now, the anomaly cancellation condition translates into the quiver language
to the equality between
the number of incoming and outgoing arrows at every node
(cf.~\eref{anofree}). It here reads,
\beq
\sum_j (S_i.S_j) N_j=\sum_j I_{ij} N_j=0 \ \ \ \forall i \ .
\eeq
Indeed, the vanishing of this quantity for every gauge groups follows from
\eref{zero_sum}. Furthermore, the fractional brane cycle, $\sum_jS_jN_j$ must have zero intersection with all other cycles in order to be anomaly free.

Seiberg duality on the quiver corresponds to Picard-Lefschetz (PL)
transformations on the cycles, which is
a reordering of vanishing cycles. In particular, when
moving a cycle $S_j$ around a cycle $S_i$, $S_j$ becomes
\beq
S_j \rightarrow S_j+(S_i.S_j) S_i=S_j+ \det\left[ \begin{array}{cc} p_i
& q_i \\  p_j & q_j \end{array} \right] S_i \ ,
\eeq
while $S_i$ remains unchanged. This is the dualisation on node $i$.
This action can be represented by a monodromy matrix $M$ acting on the
$(p,q)$  charges of the different cycles, defined as 
\beq
(p,q)'_i=M_{ij} (p,q)_j \ .
\label{new_pq}
\eeq
The new theory has to be anomaly free, i.e., 
\beq
\sum_i N'_i (p_i,q_i)'=0, \quad
\sum_i N'_i M_{ij} (p_j,q_j)=0 \ ,
\label{new_zero_sum}
\eeq
where we have used \eref{new_pq} in the last line. 
Comparing \ref{new_zero_sum} and \ref{zero_sum}, we obtain the transformation
rule for the ranks of the gauge groups
\beq
N'_i M_{ij}=N_j \ ,
\label{transformation_ranks}
\eeq
which in vector notation reads
\beq
\vec{N'}= M^{-T} \vec{N} \ .
\label{new_ranks}
\eeq


\subsection{Alternating Dualizations}
PL monodromies are specially well suited for the computation of entire
branches of duality trees obtained
by performing alternating dualizations on two nodes of a quiver. 
The theory after any number of steps
can be determined by acting on the original one with powers of a
monodromy matrix. The results that we present
in this appendix have also been presented in \cite{fiol} (where they were
derived in the context of Weyl reflections) and in \cite{fhhi}.

\paragraph{General case}
Let us start by considering the most general case. Let us call
the nodes that will undergo alternating dualizations as 1 and
2. Furthermore, we consider 
a generic number of bifundamental fields between these two nodes,
given by the intersection  
$I_{12}=a$. Under these conditions, the matrix $M$ in \ref{new_pq}
takes the form 
\beq
M=\left( \begin{array}{cc|ccc}
a & 1 & \ 0 \ & \ 0 \  & \ldots \\
-1 & 0 & 0 & 0 & \ldots \\
\hline
0 & 0 & 1 & 0 & \ldots \\
0 & 0 & 0 & 1 & \\
\vdots & \vdots & & & \ddots
\end{array} \right) \ .
\eeq
The theory that results after $k$ dualizations will be computed using
\beq
M^k=\left( \begin{array}{c|c}
A_k & 0 \\
\hline
0 & 1
\end{array} \right) \ ,
\eeq
which has a simple block structure. The non-trivial block $A_n$ can be
calculated in closed form to be
\beq
A_k=\left( \begin{array}{ccc}
{(\lambda_+^{k+1}-\lambda_-^{k+1}) \over  (\lambda_+-\lambda_-)} & \ \
&{(\lambda_+^k-\lambda_-^k) \over (\lambda_+-\lambda_-)} \\ 
\\
-{(\lambda_+^k-\lambda_-^k) \over (\lambda_+-\lambda_-)} & \ \
&{(\lambda_+ \lambda_-^k-\lambda_- \lambda_+^k) \over
(\lambda_+-\lambda_-)}
\end{array} \right) \ ,
\qquad
\lambda_\pm := {a \pm \sqrt{a^2-4} \over 2} \ .
\eeq

\begin{figure}[ht]
  \epsfxsize = 3.5cm
  \centerline{\epsfbox{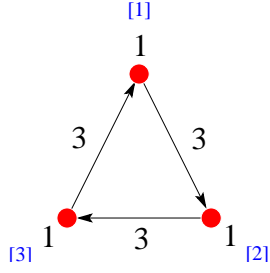}}
  \caption{Starting $dP_0$ for our alternating dualization sequence. We
  have indicated in blue the labelling of the gauge groups.} 
  \label{quiver_dP0_1}
\end{figure}

\begin{figure}[ht]
  \epsfxsize = 2.5cm
  \centerline{\epsfbox{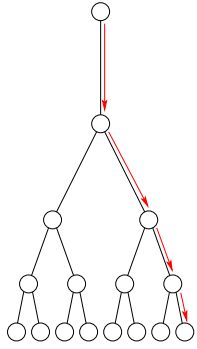}}
  \caption{Path on the duality tree when flowing to the UV with
  alternating dualizations.}
  \label{alternate}
\end{figure}

Using
the transpose of $A_k$ and \eref{new_ranks}, we have, after $k$
dualisations, the resulting ranks of the gauge group factors are
\beq
N_2(k)={(\lambda_+^{k+1}-\lambda_-^{k+1}) \over
(\lambda_+-\lambda_-)} N_2(0)-{(\lambda_+^k-\lambda_-^k) \over
(\lambda_+-\lambda_-)} N_1(0) \ ,
\label{ranks}
\eeq
with
\beq
N_1(k)=N_2(k-1) \ .
\eeq
Equation \eref{ranks} can also be used to estimate the asymptotic
behavior of the ranks for a large number of iterations. 
In the case of $a \geq 3$ we have $\lambda_+>1$ and
$\lambda_-<1$. Then, as $k \rightarrow \infty$,  
$({\lambda_- \over \lambda_+})^k\rightarrow 0$ and we have
\beq
\label{accupoint}
N_2(k) \rightarrow \frac{\lambda_+^k}{\lambda_+-\lambda_-} N_2(0) \ .
\eeq
\paragraph{An Explicit Example: $dP_0$}
Let us consider the specific case of $dP_0$, starting from the quiver
in \fref{quiver_dP0_1} (a special case of \fref{quiver_dP0})
and performing alternating dualizations
on nodes 1 and 2.
The monodromy matrices are
\begin{eqnarray}
M=\left( \begin{array}{ccc} 3 & 1 & 0 \\ -1 & 0 & 0 \\ 0 & 0 & 1
\end{array} \right)  \ , & \ \ \ \ \ & 
M^{-T}=\left( \begin{array}{ccc} 0 & 1 & 0 \\ -1 & 3 & 0 \\ 0 & 0 & 1
\end{array} \right) \ ;
\end{eqnarray}
thus, after $k$ iterations we have
\beq
\left(M^{-T} \right)^k=\left( \begin{array}{ccccc} {\left(
{3-\sqrt{5}\over 2} \right)^k \left( {3+\sqrt{5}\over 2} \right) -
\left( {3-\sqrt{5}\over 2} \right) \left( {3+\sqrt{5}\over 2}
\right)^k \over \sqrt{5}} & \ \ \ & 
-{\left( {3-\sqrt{5}\over 2} \right)^k - \left( {3+\sqrt{5}\over 2}
\right)^k \over \sqrt{5}} & \ \ \ & 0 \\ 
\\
{\left( {3-\sqrt{5}\over 2} \right)^k - \left( {3+\sqrt{5}\over 2}
\right)^k \over \sqrt{5}} & & 
{\left( {3+\sqrt{5}\over 2} \right)^{k+1} - \left( {-3+\sqrt{5}\over
2} \right)^{k+1} \over \sqrt{5}} & & 0 \\ 
\\
0 & & 0 & & 1
\end{array} \right) \ .
\eeq
Therefore, the resulting ranks of the gauge groups become
\beq
\vec{N}(k)=(N_1(k),N_1(k+1),1) \ ,
\label{rank1}
\eeq
with
\beq
N_1(k)={1\over 5 \ 2^{k+1}} \left[ (5+\sqrt{5}) (3-\sqrt{5})^k +
(5-\sqrt{5}) (3+\sqrt{5})^k \right] \ .
\label{rank2}
\eeq
From \eref{rank1} and \eref{rank2}, it is immediate to check that the
ranks satisfy the Diophantine 
equation for $dP_0$ given in \eref{diophantine}, viz., $x^2+y^2+z^2=3xyz$.
We show in \fref{alternate} the specific path along the duality tree
that corresponds to this chain of dualizations. 
This procedure can be repeated for alternating dualization chains with
other quivers as the starting point. 
%







\section{Appendix 2: Computing Horizon Volumes}
In this Appendix, we present another application of our analyses,
namely to the computation of horizon volumes.
From the AdS/CFT correspondence for 4-d SCFT's dual to geometries of
the form $AdS_5\times X^5$, one can see that the central charge  $a$
of the 4-d theory is related to the volume of $X^5$ by
(cf.~e.g.~\cite{peter})
\be
a=\frac{\pi^3}{4\mbox{Vol}(X^5)} \ .
\label{avol}
\ee

The method for computing anomalous dimensions presented in
\sref{section_computing_gammas} makes it possible 
to use the gauge theories to compute the volumes of a wide range of
non-spherical $X^5$ geometries. We
have already indirectly made this calculations for the $X^5$
associated to the conifold  
and the complex cone over the zeroth Hirzebruch surface.
In this Appendix we will extend
the discussion to the cases of complex cones over del Pezzo surfaces
(studied from the gauge theory perspective in e.g.~\cite{toric})
and  $S^5/\IZ_n$  (the result for this case is easy to guess, it is
just the volume of $S^5$ divided by $n$).
We now give the computation of the central charges for the first six Del Pezzo 
surfaces and for the orbifolds.
These results were obtained independently by us but appeared
also in two recent papers (\cite{Intri2,Herzog}). We include
them here for completeness. Horizon volume computations have also been
presented in \cite{Bergman}.

 \subsection{Orbifolds}
 The orbifolds $\IC^3/\IZ_n$ have gravity duals with geometry $AdS_5\times 
 S^5/\IZ_n$. We label the matter 
 fields in the quiver and assign anomalous dimensions as follows ($i$ 
 labels the nodes, mod($n$)):
 \beq
 X_{i,i+2} \rightarrow \gamma_X, \quad
 Y_{i,i+2} \rightarrow \gamma_Y, \quad
 Z_{i,i+1} \rightarrow \gamma_Z  \ .
 \eeq
 The superpotential has $2 n$ cubic terms all of which have the same 
 gauge coupling because of symmetry.
 The zero beta function condition reads:
 \beq
 \gamma_X+\gamma_Y+\gamma_Z=0 \ ,
\label{orbifold}
 \eeq
giving
 \beq
 a_{\mathbb{Z}_n}={n \over 96} 
 [24+(\gamma_X^3+\gamma_Y^3+\gamma_Z^3)-3(\gamma_X^2+\gamma_Y^2+\gamma_Z^2)] \  .
 \eeq

 After taking (\ref{orbifold}) into account, we have
 \beq
 a_{\mathbb{Z}_n}={n \over
 32}(2+\gamma_X)(2+\gamma_Y)(2-\gamma_X-\gamma_Y) \ .
 \eeq
 There are four critical points for $a$: $(\gamma_X$,$\gamma_Y)$
 equaling to
 $(-2,-2)$, $(-2,4)$, $(4,-2)$ and $(0,0)$, but only the last one 
 corresponds to a maximum. Therefore
 \beq
 \gamma_X=\gamma_Y=\gamma_Z=0 \quad
\Rightarrow \quad
 a_{\mathbb{Z}_n}={n \over 4 } \ .
 \eeq
 Using (\ref{avol}) we have:
 \be
 \mbox{Vol}(S^5/\IZ_n)=\frac{\pi^3}{n} \ .
\label{genord}
 \ee
 as expected.
 \subsection{Del Pezzo surfaces}
 The quivers and superpotentials for the toric theories are taken from 
 \cite{symmetries} and for the 
 non-toric ones from \cite{wijnholt}.
 \paragraph{del Pezzo 0:}
Our first example is $dP_0$, the cone over the zeroth del Pezzo surface,
 which, as mentioned before,
 is actually the orbifold $\mathbb{C}^3/\mathbb{Z}_3$.
 The anomalous dimensions are
 uniquely determined from the vanishing of the beta functions for the gauge 
 and superpotential couplings, and 
 are $\gamma_i=0$. From this we determine the central charge to be:
 \beq
 a_{dP_0}={3 \over 4} \ .
 \eeq
 This is consistent with the general result \eref{genord} for the
 Abelian orbifolds.

 \paragraph{del Pezzo 1:}
For this and subsequent del Pezzo examples we refer the reader to
 \cite{hi,toric} for their quivers and superpotentials.
 The symmetries of the $dP_1$ quiver and superpotential are the
 simultaneous exchange of $(2,3)
 \leftrightarrow (1,4)$ and charge 
 conjugation (reversal of the direction of the arrows). The SCFT
 condition on the beta-functions leaves one 
 free parameter taken here to be $\gamma_{24}$. The central charge
 takes the form 
 \beq
 a=-{3 \over 4}(2 \gamma_{24}^2-\gamma_{24}-1) \ .
 \eeq
 Maximizing a gives $\gamma_{24}=\frac{1}{4}$ and
 \beq
 \gamma_{12}=-\frac{5}{4}, \quad
 \gamma_{23}=\gamma_{41}=-\frac{1}{2}, \quad
 \gamma_{24}=\gamma_{13}=\gamma_{34}=\frac{1}{4} \ .
 \eeq
 The result for $a$ is
 \beq
 a_{dP_1}={27 \over 32} \ .
 \eeq
 \paragraph{del Pezzo 2:}
 $dP_2$ has two toric phases and they must of course have the same
 central charge, as we now demonstrate by explicit calculation.

\underline{Model I:}
The theory is invariant under exchange of nodes $1\leftrightarrow 2$.
This 
 makes
 \beq
 \begin{array}{ccc}
 \gamma_{14}=\gamma_{42}, \ \ \ &  \ \ \ \gamma_{15}=\gamma_{25}, 
	\ \ \ & \ \ 
 \ \gamma_{31}=\gamma_{32} \ .
 \end{array}
 \eeq
 Putting the beta functions to zero again leaves one of the $\gamma$'s 
 free, giving 
 \beq
 a=-{3 \over 32} (7 \gamma_{45}^2-20\gamma_{45}+4) \ .
 \eeq
 After maximizing, $\gamma_{45}=\frac{10}{7}$  and
 \beq
 a_{dP_2}={27 \over 28} \ .
 \eeq
 The other anomalous dimensions are 
 \beq
 \gamma_{14}=\gamma_{34}=-\frac{8}{7}, \quad
 \gamma_{31}=\gamma_{15}=-\frac{2}{7}, \quad
 \gamma_{35}=\frac{4}{7} \ .
 \eeq

\underline{Model II:}
 The symmetries of the theory in terms of the quiver representation
 are the simultaneous exchange of $(1,4) \leftrightarrow (2,5)$ and  
 charge conjugation. This means
 \beq
 \begin{array}{ccc}
 \gamma_{14}=\gamma_{52}, \ \ \ &  \ \ \ \gamma_{15}=\gamma_{42}, \ \ \
 & \ \  \ \gamma_{31}=\gamma_{23} \ .
 \end{array}
 \eeq
 Solving the beta function equations as before gives
 \beq
 a=-{3 \over 125} (7 \gamma_{45}^2+16\gamma_{45}-32) \ .
 \eeq
 After maximizing $\gamma_{45}=-\frac{8}{7}$ and, as expected, 
 \beq
 a_{dP_2}={27 \over 28} \ .
 \eeq
 The other anomalous dimensions are 
 \beq
 \gamma_{14}=\gamma_{45}=-\frac{8}{7}, \quad
 \gamma_{31}=\gamma_{15}=\gamma_{42}=-\frac{2}{7}, \quad
 \gamma_{34}=\gamma_{53}=\frac{4}{7} \ . 
 \eeq

 \paragraph{del Pezzo 3:}
 $dP_3$ has four toric phases. Let us do the computation for two of them.

 \underline{Model I:}
 Using the $D6$ symmetry \cite{BP},
we start with only two independent anomalous 
 dimensions $\gamma_{i \ i+1}$ and
 $\gamma_{i \ i+2}$. Solving for the vanishing of the beta functions, we 
 obtain
 \beq
 \begin{array}{cc}
 \gamma_{i \ i+1}=-1 \ , \ \ \ &  \ \ \ \gamma_{i \ i+2}=0 \ .
 \end{array}
 \eeq
 Here the central charge is immediately determined to be
 \beq
 a_{dP_3}={9 \over 8} \ .
 \eeq

\underline{Model III: }
 The symmetries are $(12)\leftrightarrow (34)$ which means
 \beq
 \begin{array}{cccc}
 \gamma_{15}=\gamma_{35}, \ \ & \ \ \gamma_{54}=\gamma_{52}, \ \ & \ \ 
 \gamma_{41}=\gamma_{21}=\gamma_{23}=\gamma_{43}, \ \ & \ \ 
 \gamma_{64}=\gamma_{62} \ .
 \end{array}
 \eeq
Putting the beta functions to zero we obtain
 \beq
 \begin{array}{ccc}
 \gamma_{21}=\gamma_{64}=-\frac{2}{3}, \ \ \ &  \ \ \ 
 \gamma_{15}=\gamma_{54}=-\frac{1}{3}, \ \ \ & \ \ \ \gamma_{54}=0 \ .
 \end{array}
 \eeq
This gives
 \beq
 a_{dP_3}={9 \over 8} \ .
 \eeq
 the same as for Model I, as it should.

 \paragraph{del Pezzo 4:}
 This is the first non-toric example.
 The theory is symmetric under any permutation of nodes 3 to 7. As before,
 we impose the vanishing of all beta functions from \eref{betas}, 
thus getting 
 \beq
 \gamma_{21}=0, \quad
 \gamma_{1i}=-\gamma_{i2}=\frac{8}{5} \ .
 \eeq
 This determines the central charge to be
 \beq
 a_{dP_4}={27 \over 20} \ .
 \eeq

 \paragraph{del Pezzo 5:}
This geometry, in the non-generic case where the 
blow-up points in $\IP^2$ are not in general position, 
is the same as for the conifold modded out by $\IZ_2\times\IZ_2$
(cf.~e.g.~\cite{unhiggsing,wijnholt}).
Imposing the conditions for the beta functions we get
 \beq
 \gamma_{ik}=\frac{5}{2}, \quad
 \gamma_{kj}=-\frac{3}{2}, \quad
 \gamma_{ji}=-1 \ ,
 \eeq
 where $i=1,2 \ \ , \ \ j=3,4 \  \ , \ \ k=5,6,7,8$.
 We find
 \be
 a_{dP_5}=\frac{27}{16} \ .
 \ee
 \paragraph{del Pezzo 6:}
 Here we find:
 \beq
 \gamma_{ij}=\gamma_{jk}=-1 \ , \quad
 \gamma_{ki}=2 \ , \quad
 \eeq
 where  $i=1,2,3 \ \ , \ \ j=4,5,6 \  \ , \ \ k=7,8,9$,
 and
 \be
 a_{dP_6}=\frac{9}{4} \ ,
 \ee
which, incidentally, is the same as for $\IC^3/\IZ_3\times\IZ_3$.

\paragraph{General del Pezzo:}
We observe a pattern indeed and posit that in general, for the cone
over $dP_n$, the central charge is
 \beq
 a_{dP_n}={27 \over 4 (9-n)} \ .
 \eeq
The appearance of the 9 is re-assuring because there are in all 9 del
Pezzo surfaces, with the 9-th one a pseudo-del-Pezzo surface sometimes
called half-K3, whose anticanonical divisor squares to 0 rather being ample.

Finally, the volume of the base of the complex cone over $dP_n$,
being inversely proportional to the central charge, should be
 \be
\label{VdPn}
 \mbox{Vol}_{dP_n}=\frac{\pi^3}{27}(9-n).
 \ee

\bibliographystyle{JHEP}

\end{document}